
\tolerance=10000
\input phyzzx.tex

\def\W {{\cal W}}

\Pubnum = {QMW-92-6}
\date = {November 1992}
 \pubtype={hep-th/9211113}

\titlepage
\title {\bf \W-GEOMETRY}
\author {C. M. Hull}
\address {Physics Department,
Queen Mary and Westfield College,
\break
Mile End Road, London E1 4NS, United Kingdom.}

\

\abstract
{ The geometric structure of theories with gauge fields of spins  two and
higher should involve a higher spin generalisation of Riemannian  geometry.
Such geometries are discussed and the case of  $\W_\infty$-gravity is
analysed in detail. While the gauge group for gravity in $d$ dimensions is the
diffeomorphism group of the space-time, the gauge group for a certain
$\W$-gravity  theory (which is $\W_\infty$-gravity in the case $d=2$) is the
group of symplectic diffeomorphisms of the cotangent bundle of the space-time.
Gauge transformations for $\W$-gravity gauge fields are given by requiring the
invariance of a generalised line element. Densities exist and can be
constructed from  the line element (generalising $\sqrt { \det g_{\mu \nu}}$)
only if $d=1$ or $d=2$,  so that only for $d=1,2$ can actions be constructed.
These two cases  and the corresponding $\W$-gravity actions are considered in
detail. In $d=2$, the gauge group is effectively only a subgroup of the
symplectic diffeomorphism group. Some of the constraints that arise for $d=2$
are similar  to equations arising in the study of self-dual four-dimensional
geometries and can be analysed using twistor methods, allowing contact to be
made with other formulations of $\W$-gravity. While the twistor transform for
self-dual spaces with one Killing vector reduces to a Legendre transform, that
for two Killing vectors  gives  a generalisation of the Legendre transform.}

 \endpage

\pagenumber=1

\def\M {{\cal M}}
\def\N {{\cal N}}

\def\F {{f}}

\REF\me{C.M. Hull, \pl\ {\bf 240B} (1990) 110.}
 \REF\van{K. Schoutens, A. Sevrin and
P. van Nieuwenhuizen, \pl\ {\bf 243B} (1990) 245.}
 \REF\pop{E. Bergshoeff, C.N. Pope, L.J. Romans, E.
Sezgin, X. Shen and K.S. Stelle, \pl\ {\bf 243B}
(1990) 350; E. Bergshoeff, C.N. Pope and K.S. Stelle, \pl\ (1990).}
\REF\meee{C.M. Hull, \pl\ {\bf 259B} (1991) 68.}
\REF\mee{C.M. Hull,  \np\ {\bf 353
B} (1991) 707.}
\REF\meeee{C.M. Hull, \np\ {\bf 364B}
(1991) 621; \np\ {\bf B367} (1991) 731;
 \pl\ {\bf 259B} (1991) 68.}
\REF\vann{K. Schoutens, A.
Sevrin and P. van Nieuwenhuizen, \np\ {\bf B349}
(1991) 791   and Phys.Lett. {\bf 251B} (1990) 355.}
\REF\mik{A. Mikovi\' c, \pl\ {\bf 260B} (1991) 75.}
\REF\mikoham {A. Mikovi\' c,
\pl\ {\bf 278B} (1991) 51.}
\REF\awa{M. Awada and Z. Qiu, \pl\ {\bf 245B} (1990) 85 and 359.}
\REF\wanog{C.M. Hull, \np\ {\bf B367} (1991) 731.}
\REF\wstr{C.N. Pope, L.J. Romans and K.S. Stelle, \pl\ {\bf 268B}
(1991) 167 and {\bf 269B}
(1991) 287.}
\REF\wrev{C.M. Hull, in {\it Strings and Symmetries 1991}, ed. by  N.
Berkovits et al,  World Scientific, Singapore Publishing, 1992.}
\REF\hegeom{C.M. Hull, \pl\ {\bf 269B} (1991) 257.}
\REF\sot{G. Sotkov and M. Stanishkov, \np\ {\bf B356} (1991) 439;
G. Sotkov, M. Stanishkov and C.J. Zhu, \np\ {\bf B356} (1991) 245.}
\REF\wot{A. Bilal, \pl\ {\bf 249B} (1990) 56;
A. Bilal, V.V. Fock and I.I. Kogan, \np\ {\bf B359} (1991) 635.}
\REF\wit{E. Witten, in {\it Proceedings of the Texas A\& M Superstring
Workshop,
 1990}, ed. by R. Arnowitt et al, World Scientific Publishing, Singapore,
1991.}
\REF\bers{ M. Berschadsky and H. Ooguri,
\cmp\ {\bf 126} (1989) 49.}
\REF\germat{J.-L. Gervais and Y. Matsuo, \pl\ {\bf B274} 309 (1992) and
Ecole Normale
preprint LPTENS-91-351 (1991); Y. Matsuo, \pl\ {\bf B277} 95 (1992)}
\REF\itz{P. Di Francesco, C. Itzykson and J.B. Zuber, Commun. Math. Phys.
{\bf 140} 543 (1991).}
\REF\ram{J.M. Figueroa-O'Farrill, S. Stanciu and E. Ramos, Leuven preprint
KUL-TF-92-34.}
\REF\zam{A.B. Zamolodchikov, Teor. Mat. Fiz. {\bf 65} (1985)
1205.}
\REF\fatty{V.A. Fateev and S. Lykanov, \intmp\ {\bf A3} (1988) 507.}
 \REF\bil{A. Bilal and J.-L. Gervais, \pl\ {\bf
206B} (1988) 412; \np\ {\bf B314} (1989) 646; \np\ {\bf B318} (1989) 579;
Ecole Normale preprint LPTENS 88/34.}
 \REF\winf{I. Bakas, \pl\ {\bf B228} (1989) 57.}
\REF\romans{L.J. Romans, \np\ {\bf B352} 829.}
\REF\bow{P. Bouwknegt and K. Schoutens, CERN preprint CERN-TH.6583/92,
to appear in Physics Reports.}
 \REF\pleb{J.F. Plebanski, J. Math. Phys. {\bf 16} (1975) 2395.}
\REF\mon{T. Aubin, {\it Non-Linear Analysis on Manifolds. Monge-Ampere
Equations}, Springer Verlag, New York, 1982.}
 \REF\riemy{B. Riemann, (1892) {\it Uber die Hypothesen welche der Geometrie
zugrunde liegen}, Ges. Math. Werke, Leibzig, pp. 272-287.}
\REF\finre{H. Rund, {\it The Differential Geometry of Finsler Spaces}, Nauka,
Moscow, 1981; G.S. Asanov, {\it Finsler Geometry, Relativity and Gauge
Theories}, Reidel, Dordrecht, 1985.}
\REF\huchir{C.M. Hull, \pl\ {\bf 206B} (1988) 234.}
\REF\batalina{I.A. Batalin and G.A. Vilkovisky, \pl\ {\bf 102B} (1981) 27
and \pr\ {\bf D28} (1983) 2567.}
\REF\lin{U. Lindstrom and M. Ro\v cek, \npb 222 (1983) 285.}
\REF\hit{
N. Hitchin, A. Karlhede, U. Lindstrom and M. Ro\v cek, Commun. Math.
Phys. 108 (1987) 535. }
\REF\va{C.M. Hull, in {\it Super Field Theories}, edited by H.C.
Lee et al, Plenum, New York, 1987.}
\REF\wnprep{C.M. Hull, \lq The Geometric Structure of $\W_N$ Gravity', in
preparation.}
 \REF\strom{A.
Strominger, \cmp\ {\bf 133} (1990) 163.}

\def\np{Nucl. Phys.}
\def\npb{Nucl. Phys. B}
\def\pl{Phys. Lett.}
\def\pr{Phys. Rev.}

\def\cmp{Commun. Math. Phys.}
\def\intmp{Intern. J. Mod. Phys.}

\def\half{{\textstyle {1 \over 2}}}

\def\IC{{\bf C}}
\def\IR{\relax{\rm I\kern-.18em R}}

\def\n{{d}}
\def\dm {\partial_{\mu}}
\def\dn {\partial_{\nu}}
\def\dr {\partial_{\rho}}

\def\ffi {\phi^i}
\def\fj {\phi^j}
\def\fk {\phi^k}
\def\ix {\int\!\!d^2x\;}

\def\hmn {h_{\mu\nu}}
\def\himn {h^{\mu\nu}}

\def\dpl {\partial_+}
\def\dmi {\partial_-}
\def\dpm {\partial_\pm}
\def\dmp {\partial_\mp}

\def\gij {g_{ij}}

\def\lie  { {\cal L}  }

\chapter {Introduction}

\W-gravity is a higher-spin generalisation of gravity which   plays an
important r\^ ole in two-dimensional physics  and has led to   new
generalisations of string theory [\me -\wstr] (for a review, see [\wrev]).
 The gauge fields are the two-dimensional
metric $\hmn$ together with a (possibly infinite) number of higher-spin
gauge fields $h_{\mu \nu \dots \rho}$.
\W-gravity can be thought of as the gauge theory of local \W-algebra
symmetries in the same sense that two-dimensional gravity can be thought
of as the result of gauging the Virasoro algebra, and different
\W-algebras lead to different \W-gravities. A \W-algebra is an extended
conformal algebra containing the Virasoro algebra and is generated by a
spin-two current and a number of other currents, including some of spin
greater than two [\zam-\romans] (for a review, see [\bow]).

A matter system with \W-algebra symmetry can be coupled to \W-gravity in
such a way that the conformal symmetry is promoted to a diffeomorphism
symmetry and the whole  \W-algebra symmetry is promoted to a local gauge
symmetry. For chiral \W-algebras, the resulting coupling is always linear
in the gauge fields [\meee,\mee], but if both left and right handed
\W-algebras are gauged, the theory is non-polynomial in the gauge fields of
spin-two and higher [\me].
For the coupling to pure gravity, the key to understanding the non-linear
structure is Riemannian geometry. The spin-two gauge field is interpreted
as a Riemannian metric and the non-linear action is then easily
constructed using tensor calculus and the fundamental density, $\sqrt
{-h}$, where $h=\det (\hmn)$. This suggests that the
non-polynomial structure of \W-gravity might be best understood in terms of
some higher-spin generalisation of Riemannian geometry and the aim of this
paper is to present just such an interpretation.
The main results, which include the construction of the
full non-linear action in closed form (without using auxiliary fields),
were first summarised in [\hegeom], but here a more detailed account  will
be given and the geometry of the results will be discussed.
Other approaches to the geometry of \W-algebras and \W-gravity are given
in [\sot-\ram].

In Riemannian geometry, the line element for a manifold $M$ is given in
terms of the metric $\hmn (x)$ by
$$
ds=(\hmn d x^\mu d x ^\nu)^{1/2}
\eqn\riem$$
An equally good line element can be defined using an $n$'th rank tensor
field $h_{\mu_1 \mu_2 \dots \mu _n}$:
$$
ds=(h_{\mu_1 \mu_2 \dots \mu _n} d x^{\mu_1} d x^{\mu_2}\dots d
x^{\mu_n})^{1/n} \eqn\riemrej$$
and this can be used to construct a geometry with almost all the features
of the usual Riemannian geometry, although Pythagoras' theorem is replaced
by a relation between the $n$'th powers of lengths.\foot{The line element
\riemrej\ is invariant under the diffeomorphisms $\delta x^\mu =-k^\mu (x)$,
$\delta h_{\mu_1 \mu_2 \dots  \mu _n}=\lie _k h_{\mu_1 \mu_2 \dots  \mu
_n}$
where $\lie _k$ denotes the Lie derivative with respect to $k^\mu$.
The transformation of $h_{\mu_1 \mu_2 \dots  \mu _n}$ can be rewritten in
a suggestive way as
 $\delta h_{\mu_1 \mu_2 \dots  \mu _n}= n\nabla _{(\mu_1}k_{\mu_2 \dots
\mu _n)}$ where $k_{\mu_2 \dots  \mu _n}=k^{\mu _1}h_{\mu_1 \mu_2 \dots  \mu
_n}$ and $\nabla $ is an affine connection constructed using $h_{\mu_1
\dots  \mu _n}$.}
 In fact, the line element \riemrej\
was considered by Riemann [\riemy], but rejected in favour of the simpler
alternative \riem.

A further generalization is to consider a line element
$$
ds=N(x,dx)
\eqn\fin$$
where $N$ is some function which is required to satisfy the homogeneity
condition
$$
N(x,\lambda dx)= \lambda N(x,dx)
\eqn\finscal$$
so that   scaling a coordinate interval   scales the
length of that interval by the same amount.
This defines a Finsler geometry [\finre] and \riem\ and \riemrej\ arise with
special choices of the Finsler metric function $N$.
The length of a curve $x^\mu (t)$ is given by $\int dt N(x,\dot x)$
and
this is invariant under reparameterizations $t \rightarrow t'(t)$ as a
result of \finscal. It is possible to define Finsler geodesics, connections,
curvatures etc [\finre] and even to attempt a Finsler generalisation of
general relativity (see [\finre] and references therein).

To describe \W-gravity, it is necessary to further generalise the geometry
by adopting a general line element \fin, without imposing the Finsler
homogeneity condition \finscal. Then $N$ is a real function on the tangent
bundle $TM$   that defines the length of a tangent
vector $y ^\mu \in T_xM$ at $x \in M$ to be $|y|=N(x,y)$.
It will prove convenient to work with the
\lq metric function' $ (x,y)=N^2(x,y)$ instead of $N$.
Expanding  in $y$
$$
\F (x,y)= \dots \hmn (x) y^\mu y ^\nu + \dots +
h_{\mu_1 \mu_2 \dots \mu_n}(x) y^{\mu_1}y^{\mu_2}\dots y^{\mu_n}+ \dots
\eqn\finex$$
gives a series  of coefficients $\hmn$, $h_{\mu_1 \mu_2 \dots \mu_n},\dots
$ and the line element will be    coordinate-independent if these transform
as tensors under diffeomorphisms of $M$.
The
 gauge fields of \W-gravity will be given a geometric meaning by relating
them to such tensors.

Similarly, the inverse metric, which defines the squared length $|y|^2=\himn
y_\mu y_\nu$ of a cotangent vector $y_\mu$   can be generalised by
introducing a \lq cometric function' $ F(x^\mu , y_\nu)$ on the
cotangent bundle and defining the length of $y_\mu \in T^*_xM$  to be
given by $|y|^2= F(x^\mu , y_\nu)$.
Expanding in $y$ as in \finex\ gives
$$
F(x^\mu ,y_\mu)= \sum_n {1 \over n} h^{\mu _1 \dots \mu _n}_{(n)}(x)
y_{\mu _1 }\dots y_{\mu _n}
\eqn\cometa$$
where the coefficients $h^{\mu _1 \dots \mu _n}_{(n)}(x)$ are contravariant
tensors on $M$, so that the
\lq length' of a cotangent vector  is  coordinate independent. For many
purposes, we will find it convenient to work with a cometric function rather
than a metric function.

We shall eventually want to regard the $h^{\mu _1 \dots \mu _n}_{(n)}(x)$ as
higher spin gauge fields on $M$ with transformations of the form
$$\delta h^{\mu _1 \dots \mu _n}_{(n)}= -nh_{(2)}^{\nu (\mu_1 \mu _2}
\partial _\nu \lambda  ^{\mu _2 \dots \mu _n)}_{(n)}+\dots
\eqn\lingaga$$
plus higher order terms involving the gauge fields, where the
infinitesimal parameter
$\lambda
^{\mu _1 \dots \mu _{n-1})}_{(n)}(x)$ is a rank $n-1$ symmetric tensor.
The cometric function \cometa\ can be regarded simply as a generating
function for these gauge fields, but as we shall see, the
gauge transformations have a
natural geometric interpretation on $T^*M$.

As the homogeneity condition \finscal\ has been dropped, it is possible to
consider a much larger group of transformations than the diffeomorphisms of
$M$, ${\rm Diff}(M)$, namely the diffeomorphisms of the tangent bundle (${\rm
Diff}(TM)$) or cotangent bundle (${\rm Diff}(T^*M)$),  which in general mix
$x$ and $y$.  It is natural to demand that
  $\F $ and $ F$ be invariant  (scalar) functions on $TM$ and $T^*(M)$, \ie\
that $F '(x',y')=F (x,y)$ etc, and the transformation $F \rightarrow F'$
corresponds to  variations under which the gauge fields  $h^{\mu_1   \dots
\mu_n}$ of different spins transform into one another. These transformations
turn out to be too general, however. Roughly speaking, this is because they do
not preserve the important difference between the coordinates $x$ on the base
manifold $M$ and the fibre coordinates $y$.  More precisely, the action of
${\rm Diff} (T^*M)$ leads to  transformations of the $h^{\mu _1 \dots \mu
_n}_{(n)}(x)$ which depend on both $x$ and $y$, and so are not of the desired
form \lingaga\ of transformations of
 higher spin gauge fields on $M$ whose transformations depend on $x$ alone.

For this reason, we seek a natural subgroup of the bundle diffeomorphisms.
 For the cotangent bundle, we consider the
 symplectic diffeomorphism group  ${\rm Diff}_0(T^*M)$ consisting of the
subgroup of the diffeomorphism group that preserves the natural  symplectic
form $\Omega =dx^\mu _\Lambda dy_\mu$. We shall discover the remarkable result
that  requiring the cometric function (restricted to   certain natural
sections of the bundle) to be invariant under  symplectic diffeomorphisms
leads to a natural set of transformations for the gauge fields $h^{\mu_1
\dots \mu_n}$ that are independent of $y$. This is true for   {\it any}
dimension of $M$. A r\^ ole for ${\rm Diff}_0(T^*M)$ has been suggested
previously in the context of \W-gravity [\wit,\pop]. As will be seen, ${\rm
Diff}_0(T^*M)$ turns out to be the gauge group for one-dimensional \W-gravity,
but for the two-dimensional case this is still too big and it is necessary to
restrict further to a subgroup of
${\rm Diff}_0(T^*M)$.

In order to construct actions, we shall need some generalisation of the
density
$\sqrt {-h}$, $h=\det [\hmn ]$. To construct an action for a scalar field
$\phi$, all that is needed is a tensor density
$\tilde h^{\mu \nu}$, as the action
$$S={1\over 2}
\int d^d x \, \tilde h^{\mu \nu}
\partial _\mu \phi \partial _\nu \phi
\eqn\erterrtyrty$$
is then invariant. The tensor density   can be regarded as an independent
field, but  as $\det [\tilde h^{\mu \nu}]$ is a scalar in two dimensions,
 one can consistently impose the constraint $\det [\tilde h^{\mu \nu}]=-1$ and
this can then be solved in terms of an unconstrained metric $\hmn$ as $ \tilde
h^{\mu \nu} = \sqrt {-h}h^{\mu \nu}$, so that \erterrtyrty\ becomes the
standard minimal coupling. The quantity $\tilde F(x,y)= {1 \over 2}\tilde
h^{\mu \nu}y_\mu y_\nu$ changes by a total derivative on $M$ under an
infinitesimal diffeomorphism, $\delta \tilde F= \partial (k^\mu \tilde
F)/\partial x^\mu$ so that  $\int_M d^dx \, \tilde F$ is invariant (with
appropriate boundary conditions).

In order to construct \W-gravity actions, we shall need a
\lq cometric density'
$$
\tilde F(x^\mu ,y_\mu)= \sum_n {1 \over n} \tilde h^{\mu _1 \dots \mu
_n}_{(n)}(x) y_{\mu _1 }\dots y_{\mu _n}
\eqn\cometda$$
which transforms by a total derivative under an infinitesimal \W-gravity gauge
transformation, so that $\int_M d^dx \, \tilde F$ is \W-invariant.
 In particular, invariance under ${\rm Diff} (M)$ will require that the
 $\tilde h^{\mu _1 \dots \mu _n}_{(n)}(x)$ transform as tensor densities under
${\rm Diff} (M)$. We will show that, with the gauge group
${\rm Diff}_0(T^*M)$,
   such cometric densities do not exist for dimensions $d>2$, that they do
exist for $d=1$   and that they do not exist for $d=2$, but that there are
cometric densities in $d=2$
  for a certain subgroup {{\cal H}} of ${\rm Diff}_0(T^*M)$. This means that
\W-gravity actions of the type investigated in this paper exist only for
$d=1,2$ and that the \W-gravity gauge group in $d=1$ is ${\rm Diff}_0(T^*M)$
while that in $d=2$ is the subgroup ${{\cal H}}\subset {\rm Diff}_0(T^*M)$.
 In one dimension, we give an explicit construction of a cometric density from
a cometric, generalising the construction $ \tilde h^{\mu \nu} = \sqrt {-
h}h^{\mu \nu}$. In $d=2$, we show that the constraint that generalises $\det
[\tilde h^{\mu \nu}]=-1$ is
$$
\det \left ({\partial ^2   \tilde F (x,y)\over \partial y_\mu
\partial y_ \nu}\right)=-1
\eqn\deconcon$$
and give some evidence to support the conjecture that a cometric
density satisfying
  this constraint can be written in terms of a cometric. This is the real
Monge-Ampere equation [\mon] and is sometimes referred to as one of
Plebanski's equations [\pleb].

The plan of the paper is as follows. In section 2,  classical \W-algebras and
linearised \W-gravity will be reviewed
 and in section 3 the construction of \W-gravities involving auxiliary
variables [\van] will be reviewed. In section 4, $d$-dimensional \W-gravity and
symplectic diffeomorphisms are introduced, cometrics and cometric densities
are analysed in section 5 and actions are constructed in section 6. In section
7, the relation between the equation \deconcon\ and  self-dual geometry in
four dimensions is used to motivate a twistor-transform solution of \deconcon\
which leads to a recovery of the  auxiliary variable formulation of section 3.
In section 8, the solution of \deconcon\ that generalises the construction $
\tilde h^{\mu \nu} = \sqrt {-  h}h^{\mu \nu}$ is considered and \W-Weyl
symmetry discussed. Section 9 summarises  the results and discusses some
generalisations.

\chapter{Classical \W-Algebras and Linearised \W-Gravity}

The $\W_\infty$  algebra [\winf] (sometimes referred to as $w_\infty$) is
a Lie algebra generated by an infinite set of currents $
W^r=\{W^2,W^3,W^4,\dots\}$  where $W^r$ has spin $r$.
Expanding $W^r$ in modes $W^r_n$, the algebra can be written as
$$
[W^r_m,W^s_n]=\{ (r-1)n-(s-1)m \} W_{m+n}^{r+s}
\eqn\winfty$$
with $r ,s\ge 2$. Expanding the range of $r,s$ to include a spin-one field
$W^1$ gives the algebra $\W_{1 +\infty}$ [\pop], which is the algebra
of symplectic diffeomorphisms of the cylinder, $\IR \times S^1$. Note that
the spin-two current $W^2$ generates a Virasoro algebra (without central
charge).

This algebra can be realised as the symmetry algebra of
the non-linear
sigma-model
with action
$$
S_0={1 \over 2} \ix \gij \dm \ffi \partial ^\mu \fj
\eqn\sig$$
where the fields  $\ffi (x^\mu)$ are maps from two-dimensional flat
space-time\foot{Throughout this paper, the two-dimensional space-time or
world-sheet will be taken to have Lorentzian signature. The conversion of
formulae to the Euclidean case is straightforward and given explicitly in
[\hegeom].} (with coordinates $x^\mu$) to a suitable target space \M, which is
some $D$-dimensional manifold with coordinates $\ffi$  ($i=1,\dots , D$) and
metric $\gij (\ffi)$.  It is useful to introduce null coordinates, $x^\pm = {1
\over \sqrt 2}
 (x^0 \pm x^1)$, so that the  flat metric is $ds^2=\eta _{\mu \nu}dx^\mu
dx^\nu=2dx^+dx^-$.  Then any symmetric rank-$s$ tensor $T_{\mu _1 \dots
\mu_s}$ is traceless ($\eta ^{\mu \nu} T_{\mu \nu \rho \sigma \dots}=0$) if
$T_{+- \rho \sigma \dots}=0$, so that it has only  two non-vanishing
components, $T_{++ \dots +}$ and $T_{-- \dots -}$, which both have spin $s$,
but have helicities $s$ and $-s$ respectively.

The spin-two currents
$$
W_{(\pm 2)} =T_{\pm \pm }={1 \over 2} \gij \dpm \ffi \dpm \fj
\eqn\stress$$
are the components of the traceless stress-energy tensor $T_{\mu
\nu}$ and satisfy the conservation laws $\dmp T_{\pm \pm}=0$. They
generate the conformal transformations $  \delta \ffi = k_\mp \dpm \ffi$
where the parameters satisfy  $\dpm k_\pm=0$; these conformal transformations
are a symmetry of \sig.
 Any symmetric tensor $d^{(s)}_{i_1 \dots i_s}(\phi)$ on \M\ can be used to
define the spin-$s$ currents
$$
W_{(\pm s)}={1 \over s} d^{(s)}_{i_1 i_2\dots i_s}\dpm \phi ^{i_1}\dpm \phi
^{i_2}\dots \dpm \phi ^{i_s}
\eqn\wis$$
which are conserved if the tensor is covariantly constant, \ie\ $\dmp
W_{(\pm s)}=0$ if
$$\nabla _jd_{i_1 \dots i_s}=0\eqn\cond$$
 where $\nabla _j$ is the covariant
derivative involving the Christoffel connection for the metric
$\gij$. If \cond\ is satisfied, these currents generate the following
semi-local symmetries of $S_0$:
$$
\delta \fj = \sum _s \lambda ^{(\pm s)} d^j_{i_1 i_2\dots i_{s-1}}\dpm \phi
^{i_1}\dpm \phi ^{i_2}\dots \dpm \phi ^{i_{s-1}}
\eqn\vaarf$$
where $\lambda ^{(\pm s)}$ are parameters of helicity $\pm 1\mp s$
satisfying $\dmp \lambda ^{(\pm s)}=0$. These non-linear transformations are
higher-spin generalisations of the spin-two conformal transformation [\me].

Any set of covariantly constant symmetric tensors on \M\
 gives in this way
a set of conserved currents $W^A$, each of which generates a semi-local
symmetry of $S_0$. The symmetry algebra and corresponding current algebra
will then only close if the tensors  satisfy certain non-linear
algebraic identities [\me,\mee]. If the current algebra is non-linear (\ie\
not a Lie algebra), as will usually be the case for algebras generated by a
finite number of currents, then the corresponding symmetry algebra has
field-dependent structure functions instead of structure constants [\me].
For $\W_\infty$, $d^{(2)}_{ij}=\gij$ and
  a rank-$s$ symmetric tensor is needed
 for each $s=3,4,\dots$.
The currents close to give the algebra \winfty\ provided the tensors
satisfy the following algebraic constraint [\mee]
$$
d^{(s)}_{l(j_{1}j_{2}...j_{s-2}}{}^{i}d^{(t)}_{j_{s-1}j_{s}...j_{s+t-3})
j_{s+t-2}}{}
^{l}=d^{(s+t-2)}_{j_{1}j_{2}...j_{s+t-2}}{}^{i}
\eqn\ytjkr$$
for all $s,t$.
For flat \M, there is a solution to this
corresponding to any Jordan algebra, with $d^{(3)}$ proportional to the
structure constants of that algebra [\romans].
For $D=1$, there is a solution with $d^{(s)}_{11\dots1}=1$
for all $s$, (corresponding to the one-dimensional Jordan algebra $\IR$).
For Jordan algebras of order $N$ (\ie\ those with a norm which is an
$N$'th degree polynomial), then, as in [\pop], the algebra \lq telescopes',
\ie\ all currents $W_{\pm s}$ of spin  $s>N$ can be written as products of
the currents of spin $s \le N$ [\mee]. Then  the algebra can be regarded
as closing on the finite set of currents of spin $\le N$, giving the
non-linear  $\W_N$ algebra, which is a certain classical limit of the
$\W_N$ algebras found in [\fatty,\bil]. For example, for Jordan algebras
with cubic norm [\romans], the spin-four current
can be written locally in
terms of the spin-two current as $W_{(\pm 4)} =W_{(\pm 2)} W_{(\pm 2)} $
and all higher currents can be written in terms of $W_{(\pm 2)} ,W_{(\pm
3)} $. This leads to (two copies of) the algebra
 $\W_3$,   generated
by $W_{(\pm 2)} ,W_{(\pm
3)} $, with classical commutation relations   given by \winfty\
for $r,s \le 3$ with $W_{(\pm 4)} =(T_{\pm \pm})^2$.
This algebra is a classical limit of Zamolodchikov's quantum operator
algebra [\zam].

The chiral semi-local \W-algebra symmetry can be
promoted to a fully local one (with parameters $\lambda ^{(\pm s)}
(x^+,x^-)$ depending on both $x^+$ and $x^-$)
by coupling to gauge fields $h^{(\pm s)}$ which transform as
$$
\delta h^{(\pm s)}=\dmp \lambda ^{(\pm s)}+O(h)
\eqn\vah$$
plus terms of higher order in the gauge fields [\me,\mee].
The linearised action is then given by
adding the Noether coupling to $S_0$, giving
$$S_0+
S_1=S_0+2\ix \sum _{s=2}^\infty \left[ h^{(+s)}W_{(+s)}+h^{(-s)}W_{(-s)}
\right] \eqn\noeth$$
   which is invariant under the linearised
transformations \vaarf,\vah\ for general local parameters $\lambda$, up to
terms dependent on the gauge fields. These can be cancelled by adding terms
of higher order in the gauge fields  and the full
gauge-invariant action, which can be constructed perturbatively to any
given order in the gauge fields using the Noether method, is non-polynomial
in the gauge fields [\me].
The aim of this paper is to investigate the full non-linear structure of this
theory and give it a geometric interpretation. Note that the linearised field
equations given by varying \noeth\ with respect to the gauge fields imply the
$W_\infty$ constraints $W_{ \pm s}=0$.

Although only the bosonic realisation of $\W_\infty$ gravity will be
considered here, other realisations and other \W-algebras can be treated
in a similar way.
For bosonic realisations, choosing different sets of $d$-tensors
satisfying different algebraic constraints gives different \W-algebras
[\mee]. Similar \W-algebra realisations are obtained in many other models,
including free-fermion theories, Wess-Zumino-Witten models
and Toda field theories. In each case, the symmetry can be gauged by
coupling to an appropriate \W-gravity, with   gauge fields corresponding
to each current [\mee].
For any model with a classical \W-algebra symmetry, the chiral gauging of
  the right-handed \W-symmetry, \ie\ the coupling to   the gauge fields
$h^{(+s)}$, is given completely by   the Noether coupling \noeth\ with
$h^{(-s)}$ set equal to zero, and no higher order terms in the gauge fields
are needed [\meee,\mee].
 For models  with a $\W_\infty$ symmetry which telescopes down to a $\W_N$
symmetry, the coupling to linearised $\W_N$ gravity is obtained by setting all
the gauge fields of spin $s>N$ to zero in the coupling to  $\W_\infty$-gravity
and modifying the transformations, as in [\pop]; however, the coupling to
non-linear $\W_N$ gravity is rather more subtle [\wnprep].

\chapter {Non-Linear Gravity and \W-Gravity}

Consider first the coupling of the sigma-model \sig\ to two dimensional
gravity. The conformal invariance implies that the only components of the
stress-energy tensor are $T_{\pm \pm}=  \gij \dpm \ffi \dpm \fj$ and
the linearised Noether coupling to the spin-two gauge fields $h_{\pm \pm }$
is given by
$$
S_{lin}= {1 \over 2} \ix \left(
  \gij \dpl \ffi \dmi \fj- h_{--}T_{++}-h_{++}
T_{--} \right)
\eqn\lingrav$$
The Noether method gives the higher order terms, which can be summed to
give
$$
S_{n}=  {1 \over 2} \ix {1 \over 1-h_{--}h_{++}}
\left(
[  1+h_{--}h_{++}]  \gij \dpl \ffi \dpl \fj- h_{--}T_{++}-h_{++}
T_{--} \right)
\eqn\nonlingrav$$
This non-polynomial action is invariant under diffeomorphisms, with $\delta
\ffi =k^\mu \dm \ffi$ and $\delta h$ as in [\huchir]. Following [\van], it can
be re-written in a polynomial \lq first-order' form as
$$\eqalign{
S_a=& 2\ix \gij  \Bigl[ \pi_+^j \dmi \ffi +\pi_-^j \dpl \ffi -\pi_+^i\pi
_-^j- \half
\dpl \ffi \dmi \fj \cr &
-\half  h_{--}\pi_+^i\pi_+^j-\half h_{++}\pi^i_-\pi_-^j \Bigr]
\cr}
\eqn\stbr$$
Solving the algebraic field equations for the auxiliary fields $\pi^i_{\pm}$
and substituting the solutions into   \stbr\ gives back the action
\nonlingrav.

Although the actions \nonlingrav\ and \stbr\ give a gauge-invariant coupling
to spin-two gauge fields, they give little insight into the geometric
structure and most would prefer to use the geometric coupling
$$
S_{geom}={1 \over 2} \ix \sqrt {-g} g^{\mu \nu} \dm \ffi \dn \fj \gij
\eqn\geograv$$
to a metric $g_{\mu \nu}$. This is invariant under  diffeomorphisms  and
 under the Weyl transformation $\delta g_{\mu \nu}=\sigma (x)g
_{\mu \nu}$. Choosing the parameterization
$$
g_{\mu\nu}=\Omega\left(\matrix{2h_{++}&1+h_{++}h_{--}\cr\noalign{\smallskip}
1+h_{++}h_{--}&
2h_{--}\cr
}\right)
\eqn\two
$$
for the metric $g_{\mu \nu}$, the conformal factor $\Omega $ drops out of
 \geograv\ as a result of Weyl invariance and the action becomes
\nonlingrav, with the singularity of \nonlingrav\ at $h_{++}h_{--}=1$
corresponding to the singularity of \geograv\ when $g=\det(g_{\mu \nu })$
vanishes.

Consider now the coupling to $\W_\infty$-gravity. The linearised coupling
is given by \noeth\ and the higher-order terms can be constructed
perturbatively, but no obvious pattern emerges and no closed form
summation analogous to \nonlingrav\ appears feasible.
The approach of [\van] gives a generalisation of the action \stbr\   that
is fully invariant under local \W-symmetries [\pop,\vann,\mee].
The action is
$$\eqalign{
S=&S_0+S_n
\cr
S_0=& 2\ix \gij  \Bigl[ \pi_+^j \dmi \ffi +\pi_-^j \dpl \ffi -\pi_+^i\pi
_-^j- \half
\dpl \ffi \dmi \fj \Bigr]
\cr
S_n=&
2\ix \sum _{s=2}^\infty \left[ h^{(+s)}W_{(+s)}(\pi)+h^{(-s)}W_{(-s)}(\pi)
\right]
\cr}
\eqn\stbrpo$$
where
$$
W_{(\pm s)}(\pi)={1 \over s} d^{(s)}_{i_1 i_2\dots i_s}\pi_\pm ^{i_1}\pi_\pm
^{i_2}\dots \pi_\pm  ^{i_s}
\eqn\wfis$$
However, it follows from Galois theory that the polynomial field equations for
the auxiliary fields $\pi_\pm^i$ cannot be solved in closed form, so that the
fields $\pi^i_\pm$ cannot be eliminated. Nevertheless, these field equations
can be solved perturbatively to any given order in the gauge fields,   and the
perturbative solution can then be used to reproduce the Noether-method
perturbative action to that order in the gauge fields.

It is clearly desirable to find a geometric approach which gives a
closed-form action to all orders in the gauge fields without using
auxiliary fields. In the coupling to gravity, the Noether approach led to
two gauge fields $h_{\pm \pm}$, which could be assembled into a symmetric
tensor $h_{\mu \nu}$ satisfying $\eta^{\mu \nu}h_{\mu \nu}=0$, where
$\eta_{\mu \nu}$ is the flat metric.
In the covariant approach, all reference to the flat metric $\eta_{\mu
\nu}$ is avoided by dropping the tracelessness condition on
$h_{\mu \nu}$. The extra component of $h_{\mu \nu}$ then decouples from
the theory as a result of Weyl invariance.

For \W-gravity, for each $s$, the two gauge fields $h^{(+s)} $ and
$h^{(-s)} $ can be assembled into a symmetric tensor $h_{\mu _1 \mu _2
\dots \mu _s}$ which is traceless, $\eta^{\mu \nu }h_{\mu \nu \dots
\rho}=0$. This suggests that the covariant theory might be written in
terms of unconstrained symmetric tensor gauge fields
$h_{\mu _1 \mu _2
\dots \mu _s}$, provided that there are higher spin generalisations of the
Weyl symmetry which can be used to eliminate the traces of the gauge
fields, so that for each $s$ all but two of the components of the gauge
field decouple. An example of such a higher-spin Weyl symmetry, which was
suggested in [\mee], is
$$
\delta   h^{(s)}_{\mu _1 \mu _2
\dots \mu _s}= h_{(\mu _1 \mu _2}  \sigma ^{(s)}_{\mu_3
\dots \mu _s)}
\eqn\swey$$
where the parameter of the Weyl-transformations for a spin-$s$ gauge field
is a rank-$(s-2)$ tensor $\sigma ^{(s)}$. It will be seen later that for a
large class of models, the covariant action can indeed be written in such
a way, with a \W-Weyl symmetry which is similar to \swey, but in which the
spin-two gauge field has no preferred r\^ ole.

We shall need a covariant generalisation of the
spin-$s$ transformation \vaarf\ which does not involve
any reference to a background world-sheet metric.
A natural guess for this  is (cf [\wit,\pop])
$$
\delta \ffi = \sum _s \lambda _{(s)}^{\mu _1 \mu_2 \dots \mu _{s-1}}
d^i_{j_1 j_2 \dots j_{s-1}}
\partial _{\mu_1} \phi ^{j_1}\partial _{\mu_2}
\phi ^{j_2} \dots \partial _{\mu_{s-1}} \phi ^{j_{s-1}}
\eqn\ltra$$
However, this corresponds to too many gauge transformations, as in the
linearised theory there are only two parameters, $\lambda ^{(+s)} $ and
$\lambda ^{(-s)} $, for each spin $s$. In the linearised theory, the
transformations
\vaarf\  can be rewritten in terms of   symmetric tensors
$\lambda _{(s)}^{\mu _1 \mu_2 \dots \mu _{s-1}}$
subject to the condition
$$\eta_{\mu \nu }\lambda _{(s)}^{\mu \nu \dots
\rho}=0
\eqn\ycon$$
which implies that, for each spin $s$, the symmetric tensor
$\lambda _{(s)}^{\mu _1 \mu_2 \dots \mu _{s-1}}$
has only two non-vanishing components, $\lambda ^{(\pm s)}$.
 In the full non-linear theory, it will be seen that the
transformations can be written as in \ltra\ but with the parameters
satisfying a  non-linear generalisation of the
tracelessness condition which is independent of the flat metric and
which
reduces to \ycon\ in the linearised theory.
These constraints can be solved in terms of some unconstrained tensors
$k ^{\mu \nu \dots}$
  in such a way that all but two of the
components of the gauge parameters $k ^{\mu \nu \dots}$ drop out
of the gauge transformation.
When expressed in terms
of the unconstrained $k ^{\mu \nu \dots}$  parameters,
the symmetry is reducible, in the sense
of [\batalina].

\chapter{ Geometry, Gravity and \W-Gravity}

Instead of restricting attention to two dimensions, it is of
interest to
attempt to formulate \W-gravity in $\n$-dimensions.
 Consider, then, the
$\n$-dimensional sigma-model or $(\n -1)$-brane in which a configuration is a
map $\ffi(x^\mu)$ from an $\n$-dimensional space-time or world-volume $\N$,
with coordinates $x^\mu$, to a $D$-dimensional target-space $\M$ with
coordinates $\ffi$.
The cotangent bundle $T^*\N$ has coordinates $(x^\mu, y_\mu)$, where
$y_\mu$ are fibre coordinates.
The map $\ffi(x^\mu)$ can be used to pull-back a metric
$\gij (\phi)$ on $\M$ to an induced metric $G_{\mu \nu}(x)= \gij (\phi (x))
\dm \ffi \dn \fj$ on $\N$. This transforms as a tensor under ${\rm Diff}(\N )$
and can be used to define actions that are invariant under ${\rm Diff}(\N )$,
such as
$$
S_{Nambu-Goto}=\int d^\n x \sqrt{-\det ( G_{\mu \nu})}, \qquad
S_{cov}=\int d^\n x \sqrt{-h} (h^{\mu \nu}G_{\mu \nu}+ \mu)
\eqn\axe$$
where $h_{\mu \nu}$ is a metric on $\N$ and $\mu$ is a constant.

In a similar way, a $\W$-metric function given by
$\F (\phi, d\phi) =\gij (\phi ) d\ffi d\fj + d_{ijk} d\ffi d\fj d\fk+\dots$ on
\M\ can be pulled back to one on \N\
$$\eqalign{&
\hat \F (x,dx)= \F (\phi(x),\dm \phi dx^\mu)
\cr &=\gij(\phi(x))
\dm \ffi \dn \fj dx^\mu dx^\nu   + d_{ijk} (\phi(x))
\dm \ffi \dn \fj \dr \fk
dx^\mu dx^\nu dx^ \rho +\dots \cr}
\eqn\pull$$
This can then be used to define an action, such as
$$
S= \int d^\n x \left[ \tilde h^{\mu \nu} _{(2)}\gij \dm \ffi \dn \fj
+ \tilde h^{\mu \nu \rho}_{(3)}
d_{ijk}
\dm \ffi \dn \fj \dr \fk + \dots \right]
\eqn\egact$$
where $\tilde h^{\mu _1 \dots \mu
_n}_{(n)}$ are some   tensor densities on $\N$.
It will be useful to introduce a generating function $\tilde
F(x^\mu, y_\mu)$ for these,
 $$
\tilde F(x^\mu ,y_\mu)= \sum_n {1 \over n} \tilde h^{\mu _1 \dots \mu
_n}_{(n)}(x) y_{\mu _1 }\dots y_{\mu _n}
\eqn\cometd$$
which can be thought of as  a variant of the cometric function
and so will be referred to as a
\lq cometric density function'.
In pure gravity, the tensor density $\tilde h^{\mu \nu} _{(2)}$ can be
written in terms of a metric tensor $  h^{\mu \nu} _{(2)}$ by
$\tilde h^{\mu \nu} _{(2)}=\sqrt{-h_{(2)}}  h^{\mu \nu} _{(2)}$, suggesting
that the cometric density might
  in turn   be related in some complicated way
to some cometric function
$$
F(x^\mu ,y_\mu)= \sum_n {1 \over n} h^{\mu _1 \dots \mu _n}_{(n)}(x)
y_{\mu _1 }\dots y_{\mu _n}
\eqn\comet$$
where the coefficients $h^{\mu _1 \dots \mu _n}_{(n)}$ are tensors on $\N$.

An important  special case is that  in which $\M$ is one-dimensional, with
$g_{11}=1,d_{11\dots 1}=1,\dots$ etc. Then a real-valued
function $\phi(x)$ on $\N$ defines a section of the cotangent bundle, $y_\mu
(x)=\dm \phi$, and the lagrangian \egact\ becomes the cometric density
$\tilde F$ evaluated on the section, $\tilde F(x^\mu , \dm \phi (x))$.

If $\gij , d_{ijk},\dots$ transform as tensors under ${\rm Diff}(\M)$, then the
line element $\F (\phi, d\phi)$ is invariant under ${\rm Diff}(\M)$ and its
pull-back $\hat \F (x,dx)$ is invariant under ${\rm Diff}(\N)$, as is the
action \egact, provided that the $\tilde h^{\mu \nu   \dots}$ transform as
tensor densities.
However, much larger non-linear symmetries can be considered which transform
tensors of different rank into one another. The ${\rm Diff}(\M)$
transformation $\delta \ffi = \xi^i(\phi)$ can be generalised to a ${\rm
Diff}(T\M)$ transformation $\delta \ffi = \xi^i(\phi,d\phi)$ and the metric
function $\F (\phi,d\phi)$ will be invariant if it is a scalar function on the
tangent bundle. The pull-back $\hat \F (x,dx)$ will then be invariant under
${\rm Diff}(T\N)$. Unfortunately, this does not lead to a natural set of
transformations for the gauge fields.

In a similar way, the cometric function \comet\ can be taken to be a scalar
under  ${\rm Diff}(T^*\N)$, so that under $ x \rightarrow x'(x,y)$, $ y
\rightarrow y'(x,y)$, the cometric \comet\ is invariant, $F'(x',y')=F(x,y)$.
However, this group is not  useful as the gauge group of \W-gravity, as it has
no natural action on the gauge fields. The relation between $\W_\infty$ and
symplectic diffeomorphisms   [\winf] suggests restricting to these  and, as we
shall see, this does lead to useful results.

 The symplectic diffeomorphisms of the cotangent bundle, ${\rm
Diff}_0(T^*\N)$, preserve the two-form $dx^\mu _\Lambda dy_\mu$ and the
infinitesimal transformations take the form
 $$
\delta x^\mu = - {\partial \over \partial y_\mu } \Lambda (x,y), \qquad
\delta y_\mu =   {\partial \over \partial x^\mu } \Lambda (x,y)
\eqn\sympd$$
for some function $\Lambda$.
The transformations \sympd\ satisfy the algebra
$$
[\delta _\Lambda ,\delta _{\Lambda '}]  = \delta _{\{ \Lambda, \Lambda
' \}}
\eqn\wsif$$
where the Poisson bracket for functions $\Lambda (x,y),\Lambda '(x,y)$ on
$T^*\N$ is $$
\{ \Lambda, \Lambda
' \}= {\partial \Lambda \over \partial x^\mu}
{\partial \Lambda '\over \partial y_\mu}-
{\partial \Lambda '\over \partial x^\mu}
{\partial \Lambda \over \partial y_\mu}
\eqn\poiss$$
This symmetry algebra is isomorphic to the $\W_\infty$-algebra [\winf].
Strictly speaking, this is
the $w_\infty$ algebra if the functions $\Lambda$ are restricted to have
the Taylor expansion
 $$\Lambda (x,y)=\sum_{s=2}^\infty \lambda
_{(s)}^{\mu _1 \dots \mu _{s-1}}(x) y_{\mu_1} \dots y_{\mu_{s-1}}
\eqn\exlam$$
 on $T^*\N$, while if this sum is extended to include a spin-one
transformation with $s=1$, then
the algebra is that denoted by $w_{1+ \infty}$ in [\pop].
 A function $F(x,y)$ transforms under these
transformations as
$F(x,y) \rightarrow F(x',y')$, which implies
$$
\delta F= \delta x^\mu {\partial F
\over \partial x^\mu}+\delta y_\mu {\partial F
\over \partial y_\mu}
= \{\Lambda ,F\}
\eqn\varff$$

Consider a section $\Sigma $ of $T^*\N$ in which the
fibre coordinate $y_\mu$ is set equal to some cotangent vector field,
$y_\mu | _\Sigma=y_\mu (x)$.
On restricting functions $\Lambda(x,y)$ on $T^*\N$ to functions
$\Lambda |_\Sigma=\Lambda(x,y(x))$ on the section,
the Poisson bracket has the
 property that
$$
\{  \left. \Lambda \right|_\Sigma, \left. \Lambda ' \right|_\Sigma
  \}=
\left.\{ \Lambda , \Lambda
' \} \right| _\Sigma +  \left.
2{\partial \Lambda \over \partial y_\mu}\right| _\Sigma
\left.
{\partial \Lambda '\over \partial y_\nu} \right| _\Sigma \partial _{[\mu }
(\left. y_{\nu]}\right| _\Sigma) \eqn\popo$$
where
$$\{  \left. \Lambda \right|_\Sigma, \left. \Lambda ' \right|_\Sigma
  \} \equiv
 {\partial \left. \Lambda \right|_\Sigma  \over \partial x^\mu}
{\partial  \left. \Lambda ' \right|_\Sigma \over \partial y_\mu(x)}-
{\partial  \left. \Lambda ' \right|_\Sigma \over \partial x^\mu}
{\partial \left. \Lambda \right|_\Sigma  \over \partial y_\mu(x)}
\eqn\blib$$
 Note $\partial _\mu y_\nu=0$, so that there are no  $\partial _\mu y_\nu$
terms in $\left.\{ \Lambda , \Lambda
' \} \right| _\Sigma$,
 but $ \partial /\partial x
^\mu (y_\nu |_\Sigma)\ne 0$. For sections corresponding to vector fields of
the form $y_\mu (x)= \dm \phi$ for some function $\phi(x)$ on \N,
 $\partial _{[\mu }
y_{\nu]}=0 $ and the Poisson brackets have the important property
$\left.\{ \Lambda |_\Sigma, \Lambda \right|_\Sigma
' \}=
\left.\{ \Lambda , \Lambda
' \} \right| _\Sigma $, so that for such vector fields it will not be
necessary to differentiate between $y_\mu$ and $y_\mu | _\Sigma=y_\mu (x)$.
  Furthermore, for such vector fields the
transformation \sympd\ on $y_\mu$ is
$$\delta (\partial _\mu \phi(x))=
{\partial \over \partial x^\mu } \Lambda (x,\partial   \phi)
\eqn\yuyutr$$
and this can be consistently rewritten in terms of
a transformation of $\phi (x)$, so that
$$\delta \phi = \Lambda (x, \dm \phi)= \sum _s \lambda
_{(s)}^{\mu _1 \dots \mu _{s-1}} \partial _{\mu_1} \phi \dots \partial
_{\mu_{s-1}} \phi \eqn\symvo$$
and this induces the following transformation
on any function $F(x^\mu , \partial _\mu \phi(x))$:
$$\delta F\equiv F(x^\mu , \partial _\mu \phi(x)+\partial _\mu
\delta\phi(x))-F(x^\mu , \partial _\mu \phi(x))
={\partial F \over \partial (\partial _\mu \phi)}
\partial _\mu \Lambda = \{ \Lambda ,F\} -\delta x^\mu \partial _\mu F
\eqn\wertet$$
The last term in \wertet, $  -\delta x^\mu \partial _\mu F$, would be
cancelled if
in addition $x$ were varied as in \sympd, in which case the result \varff\
 would
be recovered.
Transformations in which the coordinates $x^\mu,y_\mu$ transform as in
\sympd\ will be referred to as passive transformations, while
transformations such as \wertet\ in which $x^\mu$ is inert but the fields
transform as in \symvo\ will be referred to as active transformations.
Both satisfy an algebra isomorphic to the symplectic diffeomorphism algebra.
 The transformation
\symvo\ is precisely the $D=1$ form of the   transformation \ltra, with
$\phi(x)$ the bosonic field, so that these transformations indeed
satisfy the
algebra \wsif, which means that
the
one-boson realisation of $\W$-symmetry has a geometric interpretation in
terms of symplectic diffeomorphisms. Note that
no natural transformations can be obtained for the fields $\phi$ under the
full diffeomorphism group of the cotangent bundle.

\chapter {Scalars and Densities}

We now turn to the search for  natural geometric  transformations for the
gauge fields that arise in \W-gravity. Before doing this, it will be
useful to review
the derivation of the transformation of the metric in ordinary gravity. In
Riemannian geometry, a central role is played by the line element $ds^2=h_{\mu
\nu}dx^\mu dx^\nu$. Under an infinitesimal passive diffeomorphism, $\delta
x^\mu =-k^\mu(x)$ and the transformation of the metric $h_{\mu \nu}$ under
diffeomorphisms can be determined by requiring that the line element $ds^2$
 be invariant, which will be the case if $ h_{\mu \nu}$ transforms
as a second
rank tensor, $\delta h_{\mu
\nu}=2\nabla_{(\mu} k_{\nu)}$. Then $\F (x)=h_{\mu \nu}y^\mu y^\nu$
is an invariant for all vector fields $y^\mu(x)$,
 \ie\ under the transformation $x^\mu \rightarrow
x'^\mu(x)$ one has $\F '(x')=\F (x)$. Equivalently, the
transformation of the inverse metric  $h^{\mu \nu}$ can be determined by
requiring the invariance
of $h^{\mu \nu} \dm \phi \dn \phi$ for all functions $\phi$, or of
$F(x)=h^{\mu \nu}y_\mu y_\nu$ for all cotangent vector fields $y_\mu (x)$.

Instead of asking for an invariant function $F(x)=F'(x')$, it is sometimes of
interest (for example in constructing actions) to seek a density $\tilde F(x)$
such that the  integral $I=\int d^d x \tilde F(x)$ is invariant, which will be
the case if  $\tilde F'(x')=|\partial x /\partial x' |\tilde F(x)$. Requiring
that  $\tilde F(x)=\tilde h^{\mu \nu}y_\mu y_\nu$ is to be such a density for
all cotangent vector fields $y_\mu (x)$ determines the transformation of
$\tilde h^{\mu \nu}$ to be that of a tensor density. So far, $\tilde
 h^{\mu \nu}$ and $ h^{\mu \nu}$ are independent; it is a remarkable fact that
given any tensor $ h^{\mu \nu}$, one can construct a tensor density by writing
$\sqrt{-h} h^{\mu \nu}$ where $h=\det  [h_{\mu \nu}]$. If $d \ne 2$, one can
invert this and obtain $ h^{\mu \nu}$ from $\tilde h^{\mu \nu}$, while if
$d=2$, one can only obtain $ h^{\mu \nu}$ up to a local Weyl transformation.
While $ h^{\mu \nu}$ is the fundamental quantity for the discussion of
geometry, it is $\tilde h^{\mu \nu}$ which is crucial for the construction of
actions; nevertheless for Riemannian geometry the two concepts are equivalent
(modulo Weyl transformations if $d=2$).

Note that instead of dealing with passive transformations under which
the coordinates $x^\mu$ transform, the above  can be formulated in terms of
 active transformations under which the coordinates remain fixed and the
fields transform. Under active transformations, we  demand that $F(x)$
transform as a scalar, $\delta F=k^\mu  \partial _ \mu F$ and that $\tilde F$
transform as a scalar density, $\delta \tilde  F=  \partial _ \mu (k^\mu
\tilde  F)$, so that the integral $I=\int d^d x \tilde F(x)$ changes by a
surface term under diffeomorphisms.

The purpose of this section is to generalise this to obtain \W-transformations
of the gauge fields $h^{\mu \nu},h^{\mu \nu \rho},\dots$ occuring in  the $y$
expansion of some $F(x,y)$ by requiring that $F$  transform in an appropriate
fashion. The first case to be considered will be that in which $F$ is a
\W-scalar, \ie\ it is invariant  under (passive) \W-transformations, and the
gauge fields  $h^{\mu \nu},h^{\mu \nu \rho},\dots$ are all tensors. The second
case will be that in which $\tilde F(x,y)$ is a \W-density, \ie\ $\tilde
F(x,y)$ changes in such a way that $\int d^d x \tilde F$ is \W-invariant,
which will give a different set of \W-transformations for the gauge fields
$\tilde h^{\mu \nu},\tilde h^{\mu \nu \rho},\dots$ occuring in  the $y$
expansion of   $\tilde F(x,y)$, which will be tensor densities. \W-densities
will be used to construct invariant actions in the following sections. We will
concentrate on the case in which the matter system is a single free boson, as
this has a natural relation to the symplectic diffeomorphisms. However, many
of the results generalise to other matter systems and we will comment further
on this in section 9.

 Consider first a cometric function $F(x^\mu ,y_\mu)$ $(\mu=1 ,\dots \n)$ with
the $y$-expansion  \comet. Invariance of $F$ under the action of ${\rm
Diff}(\N)$ (with $y_\mu$ transforming as a covariant vector) implies that the
coefficients $h^{\mu_1 \dots \mu _m}$ in \comet\ transform as contravariant
tensors under ${\rm Diff}(\N)$. Similarly, the requirement that $F$  be
invariant under general reparameterizations of $T^* \N $, \ie\ the requirement
that $F'(x',y')=F(x,y)$ under $x \rightarrow x'(x,y)$, $y \rightarrow
y'(x,y)$, can be used to obtain transformations of the coefficients $h^{\mu_1
\dots \mu _m}$. However, in general the transformations of the tensors
$h^{\mu_1 \dots \mu _m}$ obtained in this way will be $y$-dependent and this is
unsatisfactory for the application to \W-gravity.  We shall want to interpret
the cometric as a generating functional for an infinite number of gauge fields
$h_{(n)}(x)$ which are  defined on \N\ and which transform into functions of
the gauge fields, the gauge parameters and their derivatives that are
independent of $y$. This is certainly true of the gauge fields that arise in
the Noether approach and is necessary if it is to be possible to couple the
same gauge fields to other realisations of the \W-algebra. We will now show
that if we restrict our attention to the symplectic diffeomorphisms of $\N$,
then it is possible to find $y$-independent transformations for the gauge
fields $h_{(n)}$.

For any vector field $y_\mu(x)$, the variation of $F(x,y(x))$ under the
(passive) action \sympd,\exlam\
 of the  symplectic diffeomorphisms on $x$ and $y$ is given by \varff, which
can be rewritten as
$$\eqalign{
\delta F=&\sum _{m,n=2}^\infty
\left[ -{m-1 \over n} \lambda _{(m)}^{\nu (\mu _1 \dots }\dn h_{(n)}^{\dots
\mu_{m+n-2})}+ h_{(n)}^{\nu (\mu_1 \dots} \dn \lambda _{(m)}^{\dots
\mu_{m+n-2})} \right]
y_{\mu_{1}}y_{\mu_{2}}...y_{\mu_{m+n-2}}
\cr & +
2 {\partial F \over \partial y_\mu }
{\partial \Lambda \over \partial y_\nu }\partial _{[\nu } y_{\mu]}
\cr}
\eqn\sist$$
(Note that for general $y(x)$, the $x$ variation in \sympd\ induces an extra
transformation of $y(x)$, $\delta y= \delta x^\mu \partial _ \mu y$.) If
 the tensor fields $h_{(n)}(x) $ transform under ${\rm Diff}_0(T^*\N)$ in the
following $y$-independent fashion
 $$
\delta  h^{\mu _1 \dots \mu _p}_{(p)}=p\sum_{m,n}\delta_{n+m,p+2}
\left[ {m-1 \over n} \lambda _{(m)}^{\nu (\mu _1 \dots }\dn h_{(n)}^{\dots
\mu_p)}-h_{(n)}^{\nu (\mu_1 \dots} \dn \lambda _{(m)}^{\dots \mu_p)}
\right]
\eqn\sca$$
  then the cometric function
is not
quite a scalar, but transforms under ${\rm Diff}_0(T^*\N)$ as
$$
\delta F(x,y)\equiv F(x+\delta x,y+\delta y, h+\delta h)-F(x,y,h)= 2 {\partial
F \over \partial y_\mu } {\partial \Lambda \over \partial y_\nu }\partial
_{[\nu } y_{\mu]} \eqn\varf$$
If the dimension $\n$ of $\N$ is one,  the right-hand-side vanishes and $F$
is invariant, $F'(x',y')=F(x,y)$ where $F'(x,y)$ is given by replacing
$h_{(s)}$ by $h'_{(s)}=h_{(s)}+\delta h_{(s)}$ in \comet, \ie\
$F'(x,y,h_{(s)}) \equiv F (x,y,h_{(s)}+
\delta h_{(s)})-F(x,y,h_{(s)})$. For general dimension $\n$ of $\N$, the
right-hand-side vanishes for sections
 in which $y_\mu = \dm \phi$ for some $\phi$, so that $F(x^\mu,\dm
\phi)$ is invariant under the transformations \sympd -\sca,
  restricted to the section $y_\mu = \dm \phi$.
For any dimension $\n$,
this gives a realisation of an infinite group of higher-spin gauge
transformations acting on scalar fields and gauge fields.
 The  spin-two $\lambda _{(2)}$ transformations
are just the diffeomorphisms of $\N$, with $h _{(2)}^{\mu \nu}$ the
corresponding metric gauge-field, while the $\lambda _{(s)}$ transformations
give higher spin analogues for which the gauge fields are $h _{(s)}^{\mu \nu
\dots}$.

Instead of using this passive formulation in which the coordinates $x^\mu$
transform and scalars are invariant, an (equivalent) active formulation can
be used in which the coordinates $x^\mu$ are inert and the fields $\phi$
and $h_{(n)}$ transform as in \symvo,\sca. Then instead of $\delta
 F(x^\mu,\dm
\phi)=0$, one has
$$\eqalign{
\delta
 F(x^\mu,\dm
\phi)\equiv & F(x^\mu,\dm
\phi+ \dm \delta \phi,h_{(s)}+\delta h_{(s)})-F(x^\mu,\dm
\phi,h_{(s)})
\cr =&
{\partial \Lambda \over \partial y_\mu }{\partial F(x^\mu,\dm
\phi) \over \partial x^\mu}\cr}
\eqn\asfaff$$
Such an $F(x,y)$ will be referred to as a \W-scalar.

To construct invariant actions, one needs scalar densities  rather than
scalars. It is straightforward to construct densities $D(x,y)$ that can be
integrated over the whole of the cotangent bundle (\ie\ over both $x$ and
$y$) by introducing a metric $G_{MN}$ on the cotangent bundle and
contructing the fundamental density $\sqrt {\det (G_{MN} )}$. Then $\int
d^\n x d^\n y\sqrt {G}L$ is invariant under the full group of diffeomorphisms
of the cotangent bundle for any scalar $L$. However, for \W-gravity
one requires   integrals   over the base  manifold rather than ones over
the whole bundle, \ie\ integrals of the form
$$S=
\int d^\n x \tilde F(x,y(x))
\eqn\dens$$
where $y_\mu(x)$ is some vector field. In particular, for vector fields
of the form $y_\mu(x)=\dm \phi$ the integral \dens\ becomes a candidate
action for \W-gravity.
Consider, then the integral \dens\ where the \lq cometric density function'
$\tilde F(x,y)$ has the expansion \cometd. If the coefficients
$\tilde h_{(n)}$ in \cometd\ transform as tensor densities under
${\rm Diff}(\N)$, then the integral will be invariant (up to a surface term)
under diffeomorphisms.

The next step is to attempt to find transformations of the tensor
densities $\tilde h_{(n)}$ such that the integral is invariant under
\W-transformations.
For active transformations with $x^\mu$ inert and $y_\mu$ transforming as
$$
\delta y_\mu = {\partial \Lambda \over \partial x^\mu}
\eqn\sart$$
one requires transformations of $\tilde h_{(n)}$  such that
$$
\delta \tilde F = {\partial   \over \partial x^\mu}  [\Omega ^\mu (\tilde F
,\Lambda)] \eqn\weer$$
for some $\Omega ^\mu (\tilde F ,\Lambda)$ constructed from $\tilde F
,\Lambda$ and their derivatives,
so that the integral \dens\ is invariant up to a surface term.
If
$\Omega ^\mu=\tilde F k^\mu$ for some $k^\mu (x,y(x))$, then the surface
term arising from the variation of \dens\ can be   cancelled by a
 variation of $x^\mu$, $\delta x^\mu =-k^\mu$.
That is, the  change in the measure $d^\n x$ resulting from the transformation
$\delta x^\mu =-k^\mu$ of $x^\mu$
  would be cancelled by the  variation of $\tilde F$ under the passive
transformations given by \sart, $\delta x^\mu =-k^\mu$ and $\tilde F'(x',y')=
\tilde F (x,y)J$ where $J$ is the jacobian $J= |\partial x / \partial x'|$. In
particular, it would be expected that in this case $k^\mu$ would be given by
 $k^\mu =-\partial \Lambda / \partial y_\mu $, in agreement with \sympd. If
$\tilde F$ transforms as in \weer\ for some $\Omega ^\mu$, it will be
referred to as a \W-density, while in the special case in which
$\Omega ^\mu=\tilde F k^\mu$ for some $k^\mu (x,y(x))$, so that the active
viewpoint is equivalent to a  passive one, it will be referred to as
a proper \W-density.
Surprisingly, it turns out that it is only possible to contruct \W-densities
with $y$-independent transformations of the tensor densities $\tilde h_{(n)}$
in dimensions $\n=1,2$ and that these are not proper densities, as will now be
seen. This is in contrast to the case of \W-scalars, which can be constructed
in any dimension $\n$, and the case of ordinary gravity, where densities are
proper.

Consider, then, the variation of the integral \dens\ under the $y$
transformation \sart. The change in $\tilde F$  is given by
$$\eqalign{
\delta\tilde{F}&={\partial \tilde{F}\over\partial y_{\mu}}{\partial
\over\partial x^{\mu}}\Lambda
(x,y(x))=\sum^{\infty}_{m,n=2}\Bigl[
\tilde h^{\nu\mu_{1}...\mu
_{n-1}}_{(n)}\partial_{\nu}\lambda^{\mu_{n}\mu_{n+1}...\mu_{m+n-2}}_{(m)}
y_{\mu_{1}}y_{\mu_{2}}...y_{\mu_{m+n-2}}  \cr &
+(m-1)\tilde h^{\nu\mu_{1}...\mu_{n-1}}_{(n)}\lambda^{\mu_{n}\mu_{n+1}...\mu
_{m+n-3}\rho}_{(m)}y_{\mu_{1}}y_{\mu_{2}}...y_{\mu_{m+n-3}}\partial
_{\nu}y_{\rho}
\Bigr]
\cr}
\eqn\fote$$
The strategy is to attempt to write the term involving $\partial
_{\nu}y_{\rho}$ in \fote\ as a total derivative term plus a term with no
derivatives on any $y_\mu$, as such a term can be cancelled by a suitable
variation of the gauge fields $\tilde h_{(s)}$. This would leave $\tilde F$
with  the \W-density transformation rule \weer.

In one dimension, $\n=1$, \fote\ can indeed be rewritten as
$$\delta\tilde{F}={\partial   \over \partial
x }\Omega+\sum^{\infty}_{n,m=2}{y^{n+m-2}\over
n+m-2}\left\lbrack(n-1)\mathop{\tilde{h}}\nolimits_{(n)}\partial\lambda
_{(m)}-(m-1)\lambda_{(m)}\partial\mathop{\tilde{h}}\nolimits_{(n)}\right
\rbrack \eqn\rewor$$
where
$$\Omega(x,y)=\sum^{\infty}_{n,m=2}{m-1\over n+m-2}\mathop{\tilde
{h}}\nolimits_{(n)}\lambda_{(m)}y^{n+m-2}
\eqn\ryqa$$
and the one-dimensional indices $\mu,\nu,...$ have been suppressed
($h_{(p)}=h_{(p)}^{111\dots 1}$ etc).
If
the tensor densities $\tilde h_{(p)}$  transform as
$$
\delta \tilde h_{(p)}= \sum_{m,n} \delta_{m+n,p+2}
\left[ (m-1) \lambda _{(m)}\partial
\tilde h_{(n)} -(n-1)\tilde h_{(n)}\partial \lambda _{(m)}
\right]\eqn\vogg$$
then the
variation of
$\tilde h_{(p)}$ cancels the second term on the right hand side of
\rewor, so that
$$
\delta \tilde F\equiv \tilde F(x, \phi + \delta \phi, \tilde h_{(p)} +
\delta \tilde
h_{(p)}) - \tilde F(x, \phi , \tilde h_{(p)})
= \partial_x\Omega
\eqn\varftilone$$
Then the
 integral \dens\ is invariant up to   a surface term under \sart,\vogg\
$$\delta S=\int dx\ \partial_x\Omega
\eqn\feafe$$
and this will vanish with suitable boundary conditions.
Note that $\Omega$ can be rewritten   as
$$\Omega(x,y)={\bf N}^{-1}\left({\partial\Lambda
\over\partial y}\lbrack {\bf N}\tilde{F}\rbrack\right)
\eqn\omis$$
where ${\bf N}$ is the number operator ${\bf N}=y \partial/\partial y$ for
$y$, so that ${\bf N}y^s=sy^s$ and ${\bf N}^{-1}y^s=s^{-1}y^s$.
Thus with the transformation \vogg, $\tilde F$ transforms as a \W-density,
although the surprising presence of the number operator in \omis\ implies
that it is not a proper \W-density, so that the surface term variation \weer\
cannot be completely cancelled by a transformation of $x^\mu$.
 Note that  \vogg\ implies that
$\tilde h_{(s)}$ transforms as a contravariant tensor density of weight $s$
under  the one-dimensional diffeomorphisms with parameter $\lambda=\lambda
_{(2)}$,
 $$
\delta \tilde h_{(s)}= \lambda \partial \tilde h_{(s)}- (s-1)\tilde
h_{(s)} \partial \lambda
\eqn\hdenstrans$$

In one dimension, $\n=1$, a \W-density     can be related to a \W-scalar
transformation \sca\ as follows. If $\tilde h_{(n)}$ transforms as in \vogg,
then $ h_{(n)}\equiv n \tilde h_{(n+1)}$ has precisely the transformation
\sca. This means that
 the quantity
$$K(x,y,\tilde h_{(n)})\equiv {\partial \tilde F \over  \partial y}
=\sum _n \tilde h_{(n+1)}(x) y^{n }
=\sum _n  {1 \over n} h_{(n)}(x) y^{n }
\eqn\ertteer$$
is a \W-scalar. In particular, the variation \vogg\
leads to the change of $K$
$$
\delta _h K\equiv K(x,y,\tilde h_{(n)}+\delta \tilde h_{(n)})-K(x,y,
\tilde h_{(n)})
=\partial _y \Lambda \partial _x K-\partial _y K \partial _x \Lambda=
-\{ \Lambda, K\}
\eqn\tetery$$
and the transformation of $x$ and $y$  under the symplectic diffeomorphism
\sympd\ leads to a change of $K$  given by $\delta K= \{ \Lambda, K\}$, which
cancels \tetery. This gives    the important result that for any  $\n=1$
\W-density   $\tilde F$, the derivative ${\partial \tilde F \over  \partial
y}$ is a \W-scalar.

For dimensions
$\n>1$, the problem is to write the term involving $\partial _\mu y_\nu$
in  \fote\ as a surface term plus a term without any derivatives acting on
any $y$ that can be cancelled by an appropriate $\tilde h_{(s)}$ variation.
This is not possible for $\n>2$ or for $\n =2$; this is easily seen in the
special case $\tilde F= \half \eta ^{\mu \nu} y_\mu y_\nu$, when
\fote\ becomes
$$
\delta\tilde{F}=
\partial_\nu (\Lambda y^\nu)- \Lambda (\partial_\nu   y^\nu)
\eqn\fotelin$$
and there is no way to get rid of the $\partial_\nu   y^\nu$ term (where
$y^\nu= \eta^{\mu \nu} y_\mu$).

However, for the two-dimensional case, if one further imposes the constraint
that
 $$\eta _{\mu \nu} \lambda _{(m)}^{\mu \nu \rho \dots }=0\eqn\trcono$$
then (using $\partial _\mu y_\nu =\partial _\nu y_\mu$) it follows that
$$\Lambda (\partial_\nu   y^\nu)= \left(
\partial _- \left[ \sum^{\infty}_{m=2}{2 \over m} \lambda _{(m)}
^{++\dots +} (y_+)^m
\right] -
\sum^{\infty}_{m=2}{2
\over m} (\partial _-\lambda _{(m)}^{++\dots +} )(y_+)^m \right)+
(+ \leftrightarrow -)
\eqn\rytrtyryuj$$
which is of the required form of a total derivative plus a term with no
derivatives on any $y$. Thus for $\d=2$ \W-gravity linearised about this flat
background, linearised \W-densities exist only if the gauge group is
restricted to a subgroup of the symplectic diffeomorphisms in which the
parameters satisfy a constraint whose linearised form is \trcono. This is in
agreement with the discussion of linearised \W-gravity of section 2. This
suggests that in $d=2$, a \W-density might exist if the gauge group is
restricted by some constraint whose linearearised form is \trcono, and this is
indeed the case. A lengthy calculation (using the fact that in two dimensions
any tensor can be written as
 $T_{\mu\nu \rho \dots }=T_{(\mu\nu)\rho \dots }+T_{\lbrack\mu\nu\rbrack \rho
\dots }$ and the anti-symmetric part is proportional to the two-dimensional
alternating tensor\foot{The alternating tensor satisfies
$\varepsilon^{\mu\nu}\varepsilon_{\nu\rho}=\delta^{\mu}_{\ \rho}$ and
$\varepsilon^{01}=1$.} $\varepsilon_{\mu\nu}=-\varepsilon_{\nu\mu}$:
 $T_{\lbrack\mu\nu\rbrack \rho \dots }=-{1\over2}\varepsilon_{\mu\nu}T _{\rho
\dots }$, where $T_{\rho \dots }=\varepsilon ^{\alpha\beta} T_{\alpha\beta
\rho \dots }$)
 leads to the result that \fote\ can be written as
$$\eqalign{
\delta\tilde{F}&=
{\partial   \over \partial x^\mu}
 \Omega^{\mu}+X+\sum^{\infty}_{n,m=2}{1\over
m+n-2}y_{\mu_{1}}y_{\mu_{2}}...y_{\mu_{m+n-2}}
\cr &
\times \Biggl[
(n+m-2)\mathop{\tilde
{h}}\nolimits^{\nu\mu_{1}...\mu_{n-1}}_{(n)}\partial_{\nu}\lambda
^{\mu_{n}\mu_{n+1}...\mu_{m+n-2}}_{(m)}
\cr &
-{(m-1)(n-1)\over
m+n-3}\partial_{\nu}\left(\mathop{\tilde{h}}\nolimits
^{\mu_{1}...\mu_{n}}_{(n)}\lambda^{\mathop{}\nolimits\mu_{n+1}...\mu
_{m+n-2}\nu}_{(m)}\right)
\cr &
-{(m-1)(m-2)\over
m+n-3}\partial_{\nu}\left(\mathop{\tilde{h}}\nolimits
^{\nu\mu_{1}...\mu_{n-1}}_{(n)}\lambda^{\mu_{n}\mu_{n+1}...
\mu_{m+n-2}}_{(m)}\right
) \Biggr]
\cr}
\eqn\vafiti$$
where
$$\eqalign{
X=&\sum^{\infty}_{n=2}\sum^{n-2}_{m=2}
a_{ n}
\Bigl[
\varepsilon_{\nu\sigma}\varepsilon
_{\rho\tau}\mathop{\tilde{h}}\nolimits^{\mu_{1}...\mu_{n-m-2}\nu\rho
}_{(n-m)}\lambda^{\mu_{n-m-1}...\mu_{ n-4}\sigma\tau}_{(m)}
\Bigr]
\cr &
\times
\varepsilon^{\alpha\beta}\varepsilon
^{\gamma\delta}
y_{\mu_{1}\mathop{}\nolimits
}y_{\mu_{2}}...y_{\mu_{ n-4}}y_{\alpha}y_{\gamma}\partial_{\beta}y_{\delta}
\cr}
\eqn\xis$$
for certain coefficients $a_n$,
and
$$\eqalign{
\Omega^{\nu}&=\sum^{\infty}_{n,m=2}{m-1\over(m+n-2)(m+n-3)}y_{\mu
_{1}}y_{\mu_{2}}...y_{\mu_{m+n-2}}
\cr &
\times \biggl[
(n-1)\ \left(\mathop{\tilde{h}}\nolimits^{\mu_{1}...\mu
_{n}}_{(n)}\lambda^{\mathop{}\nolimits\mu_{n+1}...\mu_{m+n-2}\nu
}_{(m)}\right)
\cr &
+(m-2)\left(\mathop{\tilde{h}}\nolimits^{\nu\mu_{1}...\mu
_{n-1}}_{(n)}\lambda^{\mu_{n}\mu_{n+1}...\mu_{m+n-2}}_{(m)}\right
)\biggr]
\cr}
\eqn\rtgjio$$
Then if the tensor densities transform as
$$\eqalign{
\delta\mathop{\tilde{h}}\nolimits^{\mu_{1}\mu_{2}...\mu_{p}}_{(p)}&=\sum
_{m,n}\delta_{m+n,p+2}\biggl[(m-1)\lambda^{(\mu_{1}\mu_{2}...}_{(m)}\partial
_{\nu}\mathop{\tilde{h}}\nolimits^{...\mu_{p})\nu}_{(n)}-(n-1)\mathop{\tilde
{h}}\nolimits^{\nu(\mu_{1}\mu_{2}...}_{(n)}\partial_{\nu}\lambda^{...\mu
_{p})}_{(m)}
\cr &
+{(m-1)(n-1)\over p-1}\partial_{\nu}\left\lbrace\lambda^{\nu(\mu
_{1}\mu_{2}...}_{(m)}\mathop{\tilde{h}}\nolimits^{...\mu_{p})}_{(n)}-
\mathop{\tilde
{h}}\nolimits^{\nu(\mu_{1}\mu_{2}...}_{(n)}\lambda^{...\mu_{p})}_{(m)}\right
\rbrace\biggr]
\cr}
\eqn\denvar
$$
the integral \dens\ transforms as
$$
\delta S=
\int d^2 x \, (\partial _\mu \Omega ^\mu +X)
\eqn\erhj$$
This means that the action will be invariant up to a surface term under the
transformations \sart,\denvar\ for which the parameters $\lambda _{(m)}$
satisfy the constraint $X=0$. This constraint gives the required non-linear
generalisation of \trcono\ and will be discussed in the next section. From
\denvar, the $\tilde h_{(s)}$ transform as tensor densities under the $\lambda
_{(2)}$ transformations. Note that on restricting to one dimension, the
transformation \denvar\ reduces to \vogg.

\chapter {Covariant Actions}

Before  constructing \W-invariant actions, it will be useful to consider the
analogous problem of deriving the coupling  of  a matter system in $\n$
dimensions to gravity, without using any knowledge of geometry.
Suppose one has a   matter current $S_{\mu \nu}=S_{(\mu \nu)}$ which
transforms under diffeomorphisms as a tensor, $\delta S_{\mu \nu}=k^\rho
\dr
S_{\mu \nu} +2 S_{\rho (\mu  }\partial _{\nu )} k^\rho$, e.g. in the
sigma-model example, one has the tensor $S_{\mu \nu}=\gij \dm \ffi \dn
\fj$. Note that it would be inappropriate at this stage to subtract a trace
to obtain the usual stress tensor, as that would involve introducing a
background metric.
  The fact that
the current $S_{\mu \nu}$ transforms linearly implies that the following
action
$$
S=\int d^\n x \, \tilde h^{\mu \nu }S_{\mu \nu}
\eqn\densact$$
can be made  diffeomorphism invariant by attributing to the field $\tilde
h^{\mu \nu }(x) $ a suitable transformation law. Indeed, the action is
invariant provided $\tilde h^{\mu \nu
}(x) $ transforms as a tensor density:
$$
\delta \tilde h^{\mu \nu }=k^\rho
\dr \tilde h^{\mu \nu }
-2\tilde h^{\rho (\mu   }\dr k^{\nu)}+
\tilde h^{\mu \nu }\dr k^\rho
\eqn\dens$$
If $\n \ne 2$, one can define $  h^{\mu \nu }=\tilde h^{\mu \nu
}[\det(-\tilde h^{\mu \nu })]^{1/(2-\n )}$ and show that it
transforms as a tensor. The density can be rewritten in terms of the
tensor as
$\tilde h^{\mu \nu }=\sqrt{-h}  h^{\mu \nu }$ (where $h=\det (h_{\mu
\nu})$ and $h_{\mu \nu}$ is the inverse of $h^{\mu \nu}$) and this can be
substituted into the action \densact\ to give the usual coupling to a
metric tensor  $  h^{\mu \nu }$. Both $h$ and $\tilde h$ have the same
number of components and the two formulations are equivalent (at least for
non-degenerate metrics). If $\n=2$, however, the tensor density cannot be
written in terms of a tensor in this way. Nevertheless, in two dimensions,
$\det(   \tilde h^{\mu \nu }) $ is a scalar, so that one can consistently
impose  the constraint $\det(
\tilde h^{\mu \nu }) =-1$ to eliminate one of the three components of
$\tilde h^{\mu \nu}$.
 This constraint can then be solved in terms of an unconstrained tensor
$h^{\mu \nu }$ by writing $\tilde h^{\mu \nu }=\sqrt{-h}  h^{\mu \nu }$.
This solution is invariant under Weyl scalings of the metric, $h_{\mu \nu}
\rightarrow \sigma h_{\mu \nu}$, so that $\tilde h_{\mu \nu}$ depends
on only two of the three components of $h_{\mu \nu}$, as one of the
components is pure gauge.

To summarise, the geometric coupling to gravity was recovered by first
finding a gauge field $\tilde h$ in terms of which the action was linear
and then rewriting this in terms of a gauge field with
covariant transformation properties in the case $\n \ne 2 $, or imposing a
covariant constraint in the case $\n =2$. We now use a similar approach to
seek the coupling of a sigma-model to $\n$-dimensional \W-gravity, which in
the case $\n =2$ has the linearised form \noeth. Consider      the case
in which the target space dimension is $D=1$. We require an action
of the form
$$
S= \int d^\n x \tilde F(x^\mu , \dn \phi)
\eqn\acto$$
for some
 cometric \W-density $\tilde F$, with expansion \cometd\ in terms of the
tensor densities $\tilde h^{\mu _1 \dots \mu _n}_{(n)}$ on \N,  and demand
that it have a \W-symmetry invariance under which $\phi$ transforms as in
\symvo\ and the transformations of the density gauge fields $\tilde h^{\mu _1
\dots \mu _n}_{(n)}$ are independent of $\phi$. \foot{If this   requirement
were dropped, it would be straightforward to find a \W-gravity coupling for all
$\n$, but it would not give a universal \W-gravity which could be coupled to
all matter systems with \W-symmetry and would not give the non-linear form of
the linearised action \noeth. Note also that an active viewpoint is now
adopted, so that the coordinates $x$ do not transform.} If   such an action is
found, the next stage is to rewrite in terms of  a cometric \comet\ whose
components $h_{(n)}$ are tensors if $\n \ne 2$, or, if $\n=2$,  to impose
invariant constraints and solve in terms of a cometric function with
higher-spin \W-Weyl symmetries, so as to recover the linearised results given
earlier. Note that the gravitational coupling for any tensor current $S_{\mu
\nu}$ is given by \densact. For \W-gravity, we will find the coupling for the
matter currents $\partial_{\mu _1 }\phi\dots \partial_{\mu _n}\phi$, but the
same coupling then immediately works for any set of matter currents $ S_{\mu
_1 \dots \mu  _n}^{(n)}$ which transform into one another under
\W-diffeomorphisms in the same way as $ \partial_{\mu _1 }\phi\dots
\partial_{\mu _n}\phi$.

The results are as follows. Invariance of the action \acto\ (up to surface
terms) requires that
  $\tilde F$ is a \W-density transforming as \weer\ for some $\Omega ^\mu$,
and the results of the last section can now be applied to the cotangent vector
field $y_\mu = \dm \phi$.
 First, if $\n >2$, there are no \W-densities for which $\tilde h_{(s)}$ has
no $\phi$ dependence in its transformation rules and so there
 is no such invariant action.

Next, if $\n =1$, so that the sigma-model
can be interpreted as a particle action,
\W-densities indeed exist, so that \W-gravity actions can be constructed.
Specifically, the action
\acto,\cometd\ is invariant (up to a surface term) under the transformations
\exlam,\symvo\ and \vogg\
where the one-dimensional indices $\mu,\nu,...$ have been suppressed.
The gauge group is the symplectic diffeomorphism group  of the cotangent
bundle of the one-dimensional manifold \N, ${\rm Diff} _0 (T^*\N)$.
This gives the one-dimensional \W-gravity theory of [\mee].

In one dimension, one can in fact go much further and construct the action
explicitly from an invariant cometric line element (\ie\ a \W-scalar)
$F(x,y)$.  For comparison, the coupling to gravity (as opposed to \W-gravity)
is given by the truncation of the action \acto\ to $\half \int dx \tilde
h_{(2)} (\partial \phi)^2$ and in this case the tensor density $\tilde
h_{(2)}$ can be rewritten in terms of a contravariant inverse metric tensor $
h_{(2)}$ by
$$\tilde h_{(2)}= \sqrt{ h_{(2)} }\eqn\roro$$
(This is the one-dimensional form of
$\tilde h^{\mu \nu}=\sqrt{h}h^{\mu \nu}$, with positive
definite metric $h_{11}
=(h^{11})^{-1} =h = \det [h_{\mu \nu}]$.)
In a similar spirit, we will now show that
 the action \acto,\cometd\ with $\n =1$
can   be rewriten in terms of a cometric
$$
F(x,\partial \phi)=\sum _{n=2}^\infty {1 \over n} h_{(n)} (\partial \phi)^n
\eqn\coco$$
which transforms as a scalar under ${\rm Diff} _0 (T^*\N)$, \ie\ under the
(active) transformation in which $h_{(n)}$ transforms as in \sca\ and
$\phi$ as in \exlam,\symvo, $\delta F= [\partial \Lambda / \partial
(\partial \phi)] \partial _x F$. (Equivalently, $F$ is invariant under the
\lq passive' transformations in which, in addition to the above
transformations, the coordinate $x$ transforms as $\delta x =-[\partial
\Lambda / \partial (\partial \phi)]$.)

It was seen in the last section that, given  any
\W-density  $\tilde F$,
the derivative ${\partial   \tilde F  \over \partial y}$ is a \W-scalar.
This suggests identifying ${\partial   \tilde F  \over \partial y}$ with the
\W-scalar \coco. However, this is not quite correct, since  the Taylor
expansions of $\tilde F$ and $F$ start at order $y^2$, while that of
${\partial   \tilde F  \over \partial y}$ starts at order $y$. It follows that
$\left({\partial   \tilde F  \over \partial y}\right)^2$ is a \W-scalar whose
expansion starts at order $y^2$ and so can be identified with \coco.  We then
give the cometric density function $\tilde F$   in terms of a \W-scalar
cometric function $F$ as
$$
 \left[ {\partial  \tilde F(x,y) \over \partial y}\right] ^2=  2 {F(x,y)  },
\qquad
{\partial   \tilde F(x,y) \over \partial y} =  \sqrt {2F(x,y)  }
\eqn\fisf$$
The factor of two is a consequence of our conventions, while the
 square root in \fisf\ was
to be expected from comparison with the pure gravity limit; indeed, the term
of lowest order in $y$ in the Taylor expansion of \fisf\ reproduces \roro.
These relations can be integrated to give $\tilde F$ explicitly, using the
boundary condition that $ \tilde F(x,y) $ is a power series in $y$ starting
with the $y^2$ term. In terms of the number operator ${\bf N}=y{\partial \over
\partial y}$, using the identity $\tilde F ={\bf N} ^{-1}(y \partial \tilde F
\partial y)$, we have $$
\tilde F (x,y)= {\bf N} ^{-1} \left[ y \sqrt {2F(x,y)} \right]
\eqn\werwgr$$
so that
$$
{\bf N} \tilde F (x,y) \equiv \sum_{n=2}^{\infty} \tilde h_{(n)} y^n
=\left( 2\sum_{n=2}^{\infty}   h_{(n)} y^{n+2} \right) ^{{1 \over 2}}
\eqn\hhtif$$
and expanding in $y$ gives the $\tilde h_{(n)}$ in terms of the $h_{(n)}$.

We turn  now to the case of two-dimensional \W-gravity. Consider the action
\acto,\cometd\  and $\phi$ transformation \exlam,\symvo\ with $\n =2$. The
tensor densities $\tilde h_{(s)}$ will eventually be expressed in terms of
tensors  $  h_{(s)}$ in such a way that in the linearised limit, \vaarf,\vah\
and \noeth\ will be recovered. However, even in the linearised theory, the
action was not invariant under the full ${\rm Diff}_0(T^*\N)$ group  under
which $\phi$ transforms as \symvo, but only under the subgroup in which the
parameters satisfied a constraint whose linearised form is \ycon. This is of
course borne out by the full non-linear analysis, with the result that the
action \cometd,\acto\ is only invariant under the $\phi$ transformation
\exlam,\symvo\ together with a transformation of the $\tilde h_{(s)}$ which is
independent of $\phi$ if the parameters $\lambda _{(s)}$ satisfy a constraint
whose linearised form is \trcono. The  transformation \vafiti\ implies that an
invariant action is obtained if the
constraint $X=0$  is imposed, where $X$ is given by \xis.
The condition that $X=0$ for all $y(x)$
implies that
$$ \sum^{n-2}_{m=2}
 \varepsilon_{\nu\sigma}\varepsilon
_{\rho\tau}\mathop{\tilde{h}}\nolimits^{\nu\rho(\mu_{1}...\mu_{n-m-2}
}_{(n-m)}\lambda^{\mu_{n-m-1}...\mu_{ n-4})\sigma\tau}_{(m)}
=0
\eqn\xiszer$$
for each $n \ge 2$.
This constrain can   be rewritten
in terms of $\Lambda$ \exlam\ and $\tilde F$ as
$$
\epsilon ^{\mu \rho} \epsilon ^{\nu \sigma} {\partial ^2 \Lambda \over
\partial y_\mu \partial y_ \nu} {\partial ^2 \tilde F \over \partial y_\rho
\partial y_ \sigma}=0
\eqn\conl$$
or, equivalently,
$$
\left[ {\partial ^2 \tilde F \over \partial y_\mu
\partial y_ \nu} \right]^{-1}{\partial ^2 \Lambda \over
\partial y_\mu \partial y_ \nu} =0
\eqn\conlva$$
Introducing frames $\tilde e_\mu ^a$ such that $\tilde h^{\mu
\nu}_{(2)}=\tilde e^\mu _a \tilde e^\nu _b \eta^{ab}$ and expanding \conl\
in $y$ gives the first few constraints as
$$\eqalign{&\eta_{ab}\lambda _{(3)}^{ab}=0, \quad
\eta_{ab}\lambda _{(4)}^{abc}={2 \over 3} \tilde h_{(3)}^{abc}
\lambda _{(3) ab}, \cr &
\eta_{ab}\lambda _{(5)}^{abcd}=\tilde h_{(3)}^{ab(c}\lambda _{(4)ab}^{d)}
-\tilde h_{(3)a}^{a (c}\lambda _{(4)b}^{d)b}
+{1 \over 2}\tilde h_{(4)}^{abcd}
\lambda _{(3) ab} \cr }
\eqn\cod$$
This generalises \ycon\ and   the trace of $\lambda
_{(s)}$ is set equal to  an $\tilde h$-dependent expression  involving  the
$\lambda _{(r)}$ for $r<s$, so that these constraints can be solved in
terms of the  trace-free parts of the parameters, leaving just two
parameters for each spin.

The action \acto,\cometd\ is then invariant under the transformations
\symvo,\exlam\ and \denvar\
provided the parameters satisfy the constraint \conl.
As in the case of gravity, the linear coupling to tensor densities is
fully gauge-invariant, but is non-minimal. In the case of gravity, the
constraint $\det( \tilde h^{\mu \nu})=-1$ can be imposed and solved as
$\tilde h^{\mu \nu}= \sqrt{-g}g^{\mu \nu}$ to give the usual Weyl-invariant
formulation \geograv. For \W-gravity, some generalisation of this constraint
is needed that is preserved by \W-gravity transformations.
{}From the   analysis of section 2, the linearised form of this  constraint
should imply that the $\tilde h^{\mu \dots \nu}$ are traceless (with respect
to the flat metric $\eta_{\mu \nu}$ about which one is expanding in the
linearised approximation), but the non-linear constraint should not refer to
any fixed background metric.
Consider the
following constraints on the gauge fields $h_{(2)},h_{(3)},h_{(4)}$
$$\det\left(\mathop{\tilde{h}}\nolimits^{\mu\nu}_{(2)}\right)=-1\eqn\decona$$
$$\mathop{\tilde{h}}\nolimits_{\mu\nu}\mathop{\tilde{h}}\nolimits
^{\mu\nu\rho}_{(3)}=0 \eqn\deconb$$
$$\mathop{\tilde{h}}\nolimits_{\mu\nu}\mathop{\tilde{h}}\nolimits
^{\mu\nu\rho\sigma}_{(4)}={2\over3}\mathop{\tilde{h}}\nolimits_{\mu
\alpha}\mathop{\tilde{h}}\nolimits_{\nu\beta}\mathop{\tilde{h}}\nolimits
^{\mu\beta\rho}_{(3)}\mathop{\tilde{h}}\nolimits^{\nu\alpha\sigma
}_{(3)}\eqn\deconc$$
where
 $\mathop{\tilde{h}}\nolimits_{\mu\nu}$ is the inverse of
$\mathop{\tilde{h}}\nolimits
^{\nu\rho}_{(2)}$,
 $\mathop{\tilde{h}}\nolimits_{\mu\nu}\mathop{\tilde{h}}\nolimits
^{\nu\rho}_{(2)}=\delta^{\ \rho}_{\mu}$.
Linearising these constraints implies that, as required, $ h_{(3)}$ and
 $h_{(4)}$
are traceless with respect to $\eta _{\mu \nu}$, to lowest order  in the gauge
fields. Furthermore, it is straightforward to check that these constraints are
preserved by the transformations \denvar, so that they can be consistently
imposed on the gauge fields.
The full set of constraints are generated by the constraint
$$
\det \left (\tilde G^{\mu \nu}(x,y) \right)=-1
\eqn\decon$$
where
$$\tilde G^{\mu \nu}(x,y)={\partial ^2   \tilde F (x,y)\over \partial y_\mu
\partial y_ \nu}
\eqn\gtiis$$
Expanding the constraint \decon\ in $y_\mu$,
one finds the coefficient of $y^{n+2}$ is a non-linear constraint on
$h_{(n)}$
which (for $n>0$) sets the trace
$\mathop{\tilde{h}}\nolimits_{\mu\nu}\mathop{\tilde{h}}\nolimits
^{\mu\nu\rho\sigma}_{(n)}$ equal to a non-linear function of the
 $h_{(m)}$ for $m<n$, so that the constraint has the correct linearised limit.
The first three constraints from the expansion of \decon\ are precisely
\decona,\deconb,\deconc.
 A lengthy calculation shows that this
  infinite set of constraints on the density gauge fields
$\tilde h_{(s)}^{\mu \nu \dots}$ is preserved by the transformations
\denvar, and so can be consistently imposed on the gauge fields without
spoiling the invariance of the action. Rather than give the lengthy direct
proof of this result, we shall instead present an indirect but simple
derivation of this constraint in section 7. The equation \decon,\gtiis\
is the  real Monge-Ampere equation for a function
of the two variables $y_\mu$;
this equation is discussed in detail in [\mon], where
it the existence of solutions is established (subject to certain conditions).

The constraint \decon\ can be interpreted as follows.
Let $z_\mu$ be complex coordinates on $\IR ^4$ with real part $y_\mu$, so that
$z_\mu =y_\mu +i u_\mu$ for some $ u_\mu$.
Thus $(x^\mu, z_\mu, \bar z_\mu)$ are coordinates for a bundle
$\IC T^*\N$ which is a complexification of $T^*\N$, whose fibre at $x^\mu$
is $\IC ^2$, the complexification of the cotangent space
$T^*_x\N \simeq \IR ^2$.
Then substituting $y_\mu =
\half(z_\mu + \bar  z _ \mu)$ in $ \tilde F (x,y)$ gives a function
$$
K_x( z,\bar z)
=\tilde F(x, z+ \bar z)
\eqn\kiss$$  for each point $x$
on the base space $\N$, which can be interpreted as the Kahler potential
for the metric $$
{\partial ^2    K _x ( z,\bar z)\over \partial z_\mu
\partial \bar z_ \nu}=\tilde G_x^{\mu \nu}( y)
\eqn\kah$$
on the complexified cotangent space at $x$,
$\IC T^*_x\N \simeq \IC ^2$.
The fact that $K_x$ is independent of $u_\mu = -{i\over 2}(z_\mu - \bar z_\mu)$
implies that $K_x$ is the Kahler potential for a Kahler metric of signature
$(2,2)$ on $\IR ^4$ with two commuting holomorphic Killing vectors, $\partial /
\partial u_\mu$.
The condition  $\det \left (\tilde G_x^{\mu \nu}( y) \right)=-1$ is then the
Plebanski equation [\pleb] (or complex Monge-Ampere equation [\mon]), which
requires that the metric is Ricci-flat and so hyperkahler and this implies
that for each $x$, the corresponding curvature tensor is either self-dual or
anti-self-dual. Thus, for each $x$, $\tilde F(x,y)=K_x( z,\bar z)$ is the
Kahler potential for a hyperkahler metric on $\IR ^4$ with two commuting
(tri-) holomorphic Killing vectors and signtature (2,2). (For Euclidean
\W-gravity, with $h_{\mu \nu}$ has signature (2,0) and the internal
hyperkahler metric $\tilde G^{\mu \nu}$ has signture (4,0)). Thus $K_x(z,\bar
z)$ gives a two-parameter family of metrics labelled by the points $x^\mu \in
\N$, so that in this way we obtain a bundle over $\N$ whose fibres are $\IC
^2$, equipped with a half-flat metric with two Killing vectors.

If $\tilde F$ satisfies the constraint \decon, the constraint \conl\ on the
infinitesimal parameters $\Lambda$ can be rewritten, to lowest order in
$\Lambda$,
 as
$$
\det \left ({\partial ^2   [\tilde F +\Lambda](x,y)\over \partial y_\mu
\partial y_ \nu}  \right)=-1
\eqn\lamcon$$
which implies that $\tilde F +\Lambda$   also corresponds to a Kahler
potential for a hyperkahler metric with two killing vectors, so that
for each $x$,  $\Lambda$ represents a deformation of the hyperkahler
geometry.

The field equation obtained by varying $\phi$ in \acto\ is
$${ \partial    \over \partial  x^\mu}{ \partial    \over \partial  y_\mu}
\tilde F=0\eqn\dsfsdfg$$
which can be rewritten as
$$ \sum_n \partial _\mu \left[ \tilde h_{(n)}^{\mu \nu_1 \dots \nu _{n-1}}
\partial _{\nu_1} \phi \dots \partial _{\nu_{n-1}} \phi \right] =0
\eqn\yuhieort$$

\chapter {Twistor Transform Solution of Monge-Ampere-Plebanski Constraints}

The general solution of the Monge-Ampere equation \decon\ can be given
implicitly
 by a Penrose transform construction. For solutions with one (triholomorphic)
Killing vector, the Penrose transform reduces to a Legendre transform
solution [\hit] which was first found in the context of supersymmetric
non-linear sigma-models [\lin]. This will now be used to solve \decon;
see [\hit] for a discussion of the twistor space interpretation.
It will be convenient to introduce the notation
 $y_{0}=\zeta$, $ y_{1}=\xi$. The first step is to write
$\tilde{F}(x,\zeta,\xi)$ as the Legendre transform with respect to $\zeta$ of
some $H$, so that
$$\tilde{F}(x,\zeta,\xi)= \pi\zeta -H(x,\pi,\xi)
\eqn\leg$$
where the equation
$${\partial H\over\partial\pi}=\zeta\ \eqn\treled$$
gives $\pi$ implicitly as a function of $x,\zeta,\xi$, so that
$\pi=\pi(x,\zeta,\xi)$.
Then it is straightforward to show that
$${\partial\tilde{F}\over\partial\zeta}=\pi,\ \ \ {\partial\tilde
{F}\over\partial\xi}=-{\partial H\over\partial\xi}\eqn\ertwghh$$
and
$${\partial\pi\over\partial\zeta}=
\left({\partial^{2}H\over\partial\pi^{2}}\right)^{-1}
\ \ \ \ \ \ \ \ \ {\partial\pi\over\partial\xi
}=-\left({\partial^{2}H\over\partial\pi^{2}}\right)^{-1}{\partial
^{2}H\over\partial\pi\partial\xi}
\eqn\ywws$$
and to use these to obtain
$$\eqalign{
{\partial^{2}\tilde F\over\partial\zeta^{2}}
&=\left({\partial^{2}H\over\partial\pi^{2}}\right)^{-1}
\cr
{\partial^{2}\tilde F\over \partial \zeta \partial\xi}
&=-\left({\partial^{2}H\over\partial\pi^{2}}\right)^{-1}{\partial
^{2}H\over\partial\pi\partial\xi}
\cr
{\partial^{2}\tilde F\over   \partial\xi ^2}
&=-{\partial
^{2}H\over\partial\xi^{2}}
+\left({\partial^{2}H\over\partial\pi^{2}}\right)^{-1}
\left(
{\partial
^{2}H\over\partial\pi\partial\xi}
\right) ^2
\cr}
\eqn\fmetis$$
It then follows that
$$\det\left({\partial^{2}\tilde{F}\over\partial y_{\mu}\partial y_{\nu
}}\right)=-\left({\partial^{2}H\over\partial\pi^{2}}\right)^{-1}{\partial
^{2}H\over\partial\xi^{2}}
\eqn\ertw$$
Then the Monge-Ampere equation \decon\ will be satisfied if and only if $H$
satisfies
 $${\partial^{2}H\over\partial\pi^{2}}-{\partial^{2}H\over\partial
\xi^{2}}=0
\eqn\htyhj$$
and the general solution of this is
$$H=H_{1}(x,\pi+\xi)\ +H_{2}(x,\pi-\xi)
\eqn\hiss$$
for arbitrary functions $H_1,H_2$. Then the general solution of \decon\ is
given by the Legendre transform \leg\ of \hiss\ and
  the action \acto\ can be given in the first order form
$$S=\int d^{2}x\tilde{F}(x,y)=\int d^{2}x\ \left(\pi \dot \phi-
H_{1}(x^{\mu},\pi + \phi ^ \prime)-H_{2}(x^{\mu},\pi-\phi ^ \prime )\right)
\eqn\firs$$
where
$y_{0}=\dot\phi, y_{1}=\phi ^ \prime$.
This is essentially the canonical formulation of \W-gravity of [\mikoham].
The field equation for the auxiliary field $\pi$ is \treled\ and this can be
used in principle to eliminate $\pi$ from the action. However, it will not be
possible to solve the equation \treled\   explicitly   in general.

The close relation between the forms of the action \firs\ and \stbrpo,\wfis\
suggests that there may be a covariant Legendre-type transform technique that
leads to the form of the action \stbrpo,\wfis. Indeed,   $\tilde F$ can be
written as a transform of a function $H$ as follows:
 $$\tilde{F}(x
^{\mu},y_{\nu})=2\pi^{\mu}y_{\mu}-{1\over2}\eta^{\mu
\nu}y_{\mu}y_{\nu}-2H(x,\pi)
\eqn\retywt$$
where the equation
$$y_{\mu}={\partial H\over\partial\pi^{\mu}}
\eqn\rtysd$$
implicitly determines
$\pi_{\mu}=\pi_{\mu}(x^{\nu},y_{\rho})$.
$H$ is not quite  a Legendre transform of $\tilde F$ with respect to $y_0$ and
$y_1$ because of  the $y^2$
term in \retywt.
Then
 $${\partial\tilde{F}\over\partial
y_{\mu}}=2\pi^{\mu}-\eta ^{\mu \nu}y_{\nu}\eqn\rtytyer$$ and
the transform \retywt\ can be inverted to give
$$H(x,\pi)=
- {1 \over 2}
\tilde{F}(x
^{\mu},y_{\nu})+\pi^{\mu}y_{\mu}-{1\over 4}\eta^{\mu
\nu}y_{\mu}y_{\nu}
\eqn\inve$$
where \rtytyer\ implicitly gives $y_\mu = y_\mu (x, \pi)$.
 As the transform is invertible, any $\tilde F$ can be written as the
transform of some $H$ and vice versa.
Using
$${\partial^{2}\tilde{F}\over\partial y_{\mu}\partial y_{\nu}}=-\eta
^{\mu\nu}+2\left({\partial^{2}H\over\partial\pi^{\mu}\partial\pi^{\nu
}}\right)^{-1}\eqn\rtyeyvs$$
it follows that
$$\det\left({\partial^{2}\tilde{F}\over\partial y_{\mu}\partial y_{\nu
}}\right)=-1+2\Delta^{-1}\left(\eta^{\mu\nu}{\partial^{2}H\over\partial
\pi^{\mu}\partial\pi^{\nu}}-2\right)
\eqn\rtyyegdlk$$
where
$$\Delta=\det\left({\partial^{2}H\over\partial\pi^{\mu}\partial\pi
_{\nu}}\right)
\eqn\pthi$$
Then $\tilde F$ will satisfy \decon\ if and only if its transform $H$
satisfies $${1 \over 2}\eta^{\mu\nu}{\partial^{2}H\over\partial\pi^{\mu}
\partial\pi^{\nu
}}={\partial^{2}H\over\partial\pi^{+}\partial\pi^{-}}=1
\eqn\tyghas$$
The general solution of this is
$$H=\pi^{+}\pi^{-}+L(x,\pi^{+})+\tilde{L}(x,\pi^{-})\eqn\rtyzzeg$$
which can be used to give the action
$$\eqalign{
S=&
\int d^{2}x\left(2\pi^{\mu}y_{\mu}-2H(x,\pi)-{1\over2}\eta^{\mu
\nu}y_{\mu}y_{\nu}\right)
\cr =&
\int d^{2}x\left(2\pi^{\mu}y_{\mu}-\eta
_{\mu\nu}\pi^{\mu}\pi^{\nu}-{1\over2}\eta^{\mu\nu}y_{\mu}y_{\nu}-2L(x,\pi
^{+})-2\tilde{L}(x,\pi^{-})\right)
\cr}
\eqn\grhjgdjkgs$$
The field equation for $\pi^\mu $ is \rtysd, and using this to substitute for
$\pi$ gives the action \acto\ subject to the constraint \decon.
Alternatively, expanding the functions $L,\tilde L$ as
$$\eqalign{
L(x,\pi ^+)&= \sum _{s=2}^\infty {1 \over s}h^{(+s)}(\pi ^+)^s
\cr
\tilde L(x,\pi ^-)&= \sum _{s=2}^\infty {1 \over s}h^{(-s)}(\pi ^-)^s
\cr}
\eqn\loi$$
reproduces the action \stbrpo.
In this way, the auxiliary fields $\pi^\mu$ of the approach of [\van]
have a natural twistor interpretation, and we learn that the fact that the
  actions  \firs,\stbrpo\ are linear in the gauge fields reflects the fact
that the twistor transform converts the self-duality equation into a
 linear twistor-space
problem.

Conversely, we know from [\van,\pop] that the action \stbrpo\ is invariant
under $\W _\infty$ transformations and that this action can be rewritten  as
\grhjgdjkgs\  provided that $H$ satisfies  the constraint \tyghas. However,
$H(x,\pi)$ can be expressed in terms of a function $\tilde F(x,y)$ using the
inverse transform \inve, and using this the action becomes simply $\int d^{2}x
\, \tilde F(x,y)$ while from \rtyyegdlk\ it follows that the constraint
\tyghas\ becomes precisely \decon. Thus the fact that \grhjgdjkgs\ subject to
\tyghas\ is an invariant action implies that the action \acto\ subject to the
constraint \decon\ is also invariant.  This establishes the result that the
constraint \decon\ is consistent with the \W-transformations, as claimed in
the last section.

\chapter{Covariant Formulation and \W-Weyl Invariance}

The constraints on the gauge fields $\tilde h_{(n)}^{\mu \nu \dots}$
generated
by \decon\ can be solved in terms of   unconstrained
gauge fields in a number of ways.
We shall first review the solution of [\hegeom]
which led to gauge fields which transformed naturally  under \W-Weyl
symmetry and then discuss a solution which it is conjectured will lead to an
expression of the gauge fields  $\tilde h_{(n)}^{\mu \nu \dots}$ occuring in
the expansion of a \W-density $\tilde F(x,y)$ in terms of the
gauge fields
$ h_{(n)}^{\mu \nu \dots}$ in the expansion of a \W-scalar.

 The constraint \decona\ can be solved in terms of
an unconstrained metric tensor $g_{\mu \nu}=g_{(2)\mu \nu}$ as
 $$\tilde h^{\mu \nu}_{(2)}=\sqrt {- g}g^{\mu \nu}_{(2)}
\eqn\decsolva$$
Similarly, the constraint
\deconb\ can be solved in terms of an unconstrained third rank tensor
$g_{(3)}^{\mu \nu \rho}$:
$$\tilde h_{(3)}^{\mu \nu \rho}=\sqrt {- g}\left[
g_{(3)}^{\mu \nu \rho} -{3 \over 4 }g^{(\mu \nu} g_{(3)}^{  \rho)
\alpha \beta}
g_{\alpha \beta} \right]
\eqn\decsolvb$$
and \deconc\ can be solved in terms of an unconstrained fourth rank tensor
$\hat g_{(4)}^{\mu \nu \rho \sigma}$:
$$\eqalign{
\tilde h_{(4)}^{\mu \nu \rho \sigma}=&
\sqrt {- g}\bigl[
\hat g_{(4)}^{\mu \nu \rho \sigma} + g^{(\mu \nu} Q^{  \rho\sigma)}
\cr &-
{1\over 8}
g^{(\mu \nu}g^{  \rho\sigma)} Q^{
\alpha \beta  }
g_{\alpha \beta}
\bigr]
\cr}
\eqn\decsolvc$$
where
$$\eqalign{
Q^{  \rho\sigma} =&
{2\over 3}
h^{\alpha \gamma (\rho}h^{\sigma) \beta
\delta}g_{\alpha \beta} g_{\gamma \delta }
 - \hat g_{(4)}^{  \rho\sigma
\alpha \beta}
g_{\alpha \beta}
\cr
h ^{\mu \nu \rho}= &
g_{(3)}^{\mu \nu \rho} -{3 \over 4 }g^{(\mu \nu} g_{(3)}^{  \rho)
\alpha \beta}
g_{\alpha \beta}
\cr}
\eqn\decsolvcaux$$

This can be repeated for all spins, giving the constrained
tensor densities
${ {\tilde h }}
^{\mu_{1}\mu_{2}...\mu_{n}}_{(n)}$ in terms
of unconstrained tensors
${ {g}}
^{\mu_{1}\mu_{2}...\mu_{n}}_{(n)}$, which can be assembled into
a function
$$ {f}(x,y)=\sum^{\infty}_{n=2}{1\over n}
{ {g}}
^{\mu_{1}\mu_{2}...\mu_{n}}_{(n)}\mathop{(x)y}\nolimits_{\mu_{1}}y_{\mu
_{2}}...y_{\mu_{n}} \eqn\fist$$
The generating function  $\tilde F$ for the tensor densities
${ {\tilde h }} _{(n)}$
can then be written as
$$\tilde F(x,y)= \Omega (x,y) f(x,y)
\eqn\rerer$$
where
$\Omega $ is determined in terms of $f$
by requiring \rerer\ to satisfy \decon.
The function $\Omega $ has an expansion of the form
$$ {\Omega}(x,y)=\sum^{\infty}_{n=0} { \Omega}
^{\mu_{1}\mu_{2}...\mu_{n}}_{(n+2)}\mathop{(x)y}\nolimits_{\mu_{1}}y_{\mu
_{2}}...y_{\mu_{n}}
\eqn\omis$$
and substituting \rerer\ in \decon\ gives a set of equations which can be
solved  to give the tensors $\Omega_{(n)}$ in terms of the unconstrained
tensors $g_{(n)}$ in \fist.
This gives
$$\eqalign{\Omega _{(2)}=& \sqrt {- g}
\cr
\Omega _{(3)}^\mu=& -{1 \over 2}
\sqrt {- g}g_{(3)}^{\mu \nu \rho}g_{\nu
\rho}
\cr
\Omega _{(4)}^{\mu \nu}=&
\sqrt {- g}\left[{1 \over 2}Q^{\mu \nu}-{1 \over 16}
g^{\mu \nu}Q^{\rho \sigma}g_{\rho \sigma} \right]
\cr}
\eqn\yioriyieepu$$
where    $Q^{\mu \nu}$ is given by \decsolvcaux\ and
$  g_{(4)}^{\mu \nu \rho \sigma}$ is related to the tensor
$\hat g_{(4)}^{\mu \nu \rho \sigma}$ in \decsolvc\ by the field redefinition
 $$g_{(4)}^{\mu \nu \rho \sigma}=
\hat g_{(4)}^{\mu \nu \rho \sigma}- {4 \over 3 }
g_{(3)}^{(\mu \nu \rho}\Omega _{(3)}^{\sigma)}
\eqn\hjdhjkh
$$

The solution \decsolva\ to the constraint \decona\
is invariant under the Weyl transformation
$g_{\mu \nu} \rightarrow \sigma(x)g_{\mu \nu}$, and this suggests that
\rerer\ should be invariant
under higher spin generalisations of this.
Indeed, writing $\tilde F$ in terms of $f$ gives an
action which is invariant under the \W-Weyl transformations
$$\delta f(x,y)=\sigma(x,y)f(x,y)
\eqn\wweylf$$
Expanding
$$\sigma(x,y)=\sigma_{(2)}(x)+\sigma^{\mu}_{(3)}(x)y_{\mu}+\sigma
^{\mu\nu}_{(4)}(x)y_{\mu}y_{\nu}+...
\eqn\exsig$$
these can be written as
$$  \delta
g^{\mu_1 \dots \mu_n}_{(n)}=n
\sum_{r=2}^n
{1 \over r}g^{(\mu_1 \dots \mu_r}_{(r)}
\sigma_{(n-r+2)}^{\mu_{r+1} \dots \mu_n)}
\eqn\exweyl$$
These transformations can be used to remove all traces from the gauge
fields, leaving only   traceless gauge fields.
These \W-Weyl transformations
are similar to those given in \swey\ and have the same
linearised limit, but have the advantage that they
do not give a privileged position to the
spin-two gauge field.

The relation
\rerer\
implies that a \W-Weyl transformation can be used to set $\tilde
F(x,y)=f(x,y)$, so that $\tilde h_{(n)}^{\mu_{1}...\mu_{n}}=g_{(n)}
^{\mu_{1}...\mu_{n}}$ in this \W-Weyl gauge.
This means that in general the transformations of
 $g_{(n)}^{\mu_{1}...\mu_{n}}$
can be taken to be equal to those of $\tilde h_{(n)}^{\mu_{1}...\mu_{n}}$,
or to be related to these by a possible \W-Weyl transformation.
For example, this gives the transformation of $g_{(2)}^{\mu \nu}$
to be that of \dens, up to a Weyl transformation
$$
\delta g^{\mu \nu }=k^\rho
\dr g^{\mu \nu }
-2g^{\rho (\mu   }\dr k^{\nu)}+
g^{\mu \nu }(\sigma +\dr k^\rho)
\eqn\densqw$$
Then shifting $\sigma \rightarrow \sigma'=\sigma +\dr k^\rho$ absorbs the $
\dr k^\rho$ term into the Weyl transformation and the transformation becomes
the standard one for an inverse metric:
$$
\delta g^{\mu \nu }=k^\rho
\dr g^{\mu \nu }
-2g^{\rho (\mu   }\dr k^{\nu)}+
g^{\mu \nu } \sigma
\eqn\densqwva$$
Similarly, the term proportional to
$[\partial_{\nu}
\lambda^{\nu(\mu
_{1}\mu_{2}...}_{(p)}]\mathop{g}\nolimits^{...\mu_{p})}_{(2)}$
in the
variation $\delta\mathop{g}\nolimits^{\mu_{1}\mu_{2}...\mu_{p}}_{(p)}$
given by replacing $\tilde{h}$ by $g$ in \denvar\ can be absorbed into a
\W-Weyl transformation, but the resulting transformation for $g_{(n)}$
is not that corrsponding to that of  a \W-scalar
and does not seem to have any obvious geometric interpretation.

The constraint \lamcon\ on the parameters $\lambda _{(n)}$ can be solved in a
similar fashion in terms of unconstrained parameters $k_{(n)}^{\mu_1
...\mu_{n-1}}$ and the transformations of the unconstrained gauge fields
can be defined to take the form  $\delta { {g}}
^{\mu_{1}\mu_{2}...\mu_{n}}_{(n)}=\partial ^{(\mu_{1}}{ {k}}
^{\mu_{2}...\mu_{n})}_{(n)}+\dots$. The $g_{(n)}$ might be thought of
as gauge fields for the whole of the symplectic diffeomorphisms of $T^* \N$
(with parameters $k_{(n)}$), and appear in the action only through the
combinations $\tilde h_{(n)}$. The transformations of
$\tilde h_{(n)}$ and $\phi$ then only depend on the parameters
$k_{(n)}$ in the form $\lambda _{(n)}$.

In gravity theory,  it is sufficient to have a metric $h_{\mu \nu}$ in order
to construct actions, as densities can be constructed using $\sqrt {-{\det [h
_{\mu \nu}]}}$. In \W-gravity, it is natural to ask whether a cometric function
$F$ which transforms as a \W-scalar can be used to construct actions, and in
particular whether a \W-density $\tilde F$ can be constructed from a \W-scalar
$F$.  If so, this would lend weight to the idea that the cometric $F$ might
play a fundamental role in \W-geometry in the same way that the line element
does in Riemannian geometry.  This would be particularly attractive, as a
\W-scalar transforms naturally under the whole of the symplectic
diffeomorphisms of the cotangent bundle, ${\rm Diff}_0(T^* \N)$, while a
\W-density only transforms under the  subgroup of this defined by the
constraint \conl. Thus, as in the previous paragraph, we would have gauge
fields $h_{(n)}$ for the whole of ${\rm Diff}_0(T^* \N)$ with the  \W-scalar
transformation law
\denvar,
$$\delta { {h}}
^{\mu_{1}\mu_{2}...\mu_{n}}_{(n)}=\partial ^{(\mu_{1}}{ {k}}
^{\mu_{2}...\mu_{n})}_{(n)}+\dots
\eqn\ertrtyu$$
with $\tilde h  ^{\mu_{1} ...\mu_{n}}_{(n)}=[h  ^{\mu_{1} ...\mu_{n}}_{(n)}-
{\rm(traces)}]+ \dots$ plus non-linear terms, and
$\delta { \tilde{h}}
^{\mu_{1} ...\mu_{n}}_{(n)}=\partial ^{(\mu_{1}}{ {\lambda}}
^{\mu_{2}...\mu_{n})}_{(n)}+\dots$ where ${\lambda}
^{\mu_{1}...\mu_{n-1})}_{(n)}=[k^{\mu_{1}...\mu_{n-1})}_{(n)}-
{\rm(traces)}]+ \dots$ plus non-linear terms.

It is straightforward to show that the first few density gauge fields
$\tilde h_{(n)}$, subject to the constraints generated by
\decon,
 can be
written in terms of the first few tensor gauge fields $ h_{(n)}$ as follows:
$$\mathop{\tilde{h}}\nolimits^{\mu\nu }_{(2)}= \sqrt{-h}
h^{\mu\nu }_{(2)}\eqn\pyour$$
$$\mathop{\tilde{h}}\nolimits^{\mu\nu\rho}_{(3)}={2\over3}\sqrt{-h}\left
\lbrack
h^{\mu\nu\rho}_{(3)}-{3\over4}h^{(\mu\nu}h^{\rho)\alpha\beta
}_{(3)}h_{\alpha\beta}\right\rbrack
\eqn\reteret$$
$$\eqalign{
\mathop{\tilde{h}}\nolimits^{\mu\nu\rho\sigma}_{(4)}=&\sqrt{-h}\left
\lbrack K^{\mu\nu\rho\sigma}-h^{(\mu\nu}K^{\rho\sigma)\alpha}_{\
\ \ \ \alpha}+{1\over8}h^{(\mu\nu}h^{\rho\sigma)}K^{\alpha\beta}_{\
\ \alpha\beta}\right\rbrack
\cr &
+{2\over3\sqrt{-h}}\left\lbrack h^{(\mu\nu}\mathop{\tilde{h}}\nolimits
^{\rho}_{(3)\alpha\beta}\mathop{\tilde{h}}\nolimits^{\sigma)\alpha
\beta}_{(3)}-{1\over8}h^{(\mu\nu}h^{\rho\sigma)}\mathop{\tilde{h}}\nolimits
^{\alpha\beta\gamma}_{(3)}\mathop{\tilde{h}}\nolimits_{(3)\alpha\beta
\gamma}\right\rbrack
\cr}
\eqn\yhaerukhuwerhr$$
where indices are raised and lowered with $h^{\mu\nu }=h^{\mu\nu }_{(2)}$
and its inverse $h_{\mu\nu }$, $h=\det [h_{\mu\nu }]$ and
$$K^{\mu\nu\rho\sigma}={1\over2}h^{\mu\nu\rho\sigma}-{1\over3}h^{\alpha
(\mu\nu}_{(3)}h^{\rho\sigma)}_{(3)\alpha}
\eqn\gthhrggdfg$$
This means that given a set of gauge fields $h_{(n)}$ transforming under
${\rm Diff}_0(T^* \N)$ as in \sca, then the gauge fields $\tilde h_{(n)}$
defined by these equations transform as in \denvar. I conjecture that all the
$\tilde h_{(n)}$ can be written in terms of   $h_{(n)}$ gauge fields in this
way, although I have as yet no general proof; this is currently under
investigation.

\chapter{Summary and Discussion}

We have seen that symplectic diffeomorphisms of the cotangent bundle of the
space-time (or world-sheet) $\N$ play a fundamental role in \W-gravity,
generalising the role played by the diffeomorphisms of $\N$ in ordinary
gravity theories. For any dimension $\n$ of $\N$, we found an infinite  set of
symmetric tensor gauge fields
$h_{(n)}^{\mu_{1} ...\mu_{n}}$, $n=2,3\dots $,
 transforming under the action of a
gauge group isomorphic to ${\rm Diff}_0(T^*\N)$ as
\def\kk{{k}}
 $$
\delta  h^{\mu _1 \dots \mu _p}_{(p)}=p\sum_{m,n}\delta_{n+m,p+2}
\left[ {m-1 \over n} \kk _{(m)}^{\nu (\mu _1 \dots }\dn h_{(n)}^{\dots
\mu_p)}-h_{(n)}^{\nu (\mu_1 \dots} \dn \kk _{(m)}^{\dots \mu_p)}
\right]
\eqn\scaa$$
where $\kk _{(m)}^{\mu _1 \dots \mu _{m-1}}(x)$ are unconstrained
infinitesimal symmetric tensor parameters.
These transformations had  a  geometric interpretation: they were precisely
the transformations needed for the generating function
$$
F(x^\mu ,y_\mu)= \sum_n {1 \over n} h^{\mu _1 \dots \mu _n}_{(n)}(x)
y_{\mu _1 }\dots y_{\mu _n}
\eqn\cometa$$
to transform as a \W-scalar, \ie\ to be invariant under
the action of the gauge group ${\rm Diff}_0(T^*\N)$ (as described in section
5, with $y=\partial \phi$). This suggested regarding $F$ as the natural
generalisation of the invariant line element of Riemannian geometry.

As well as considering \W-scalars, we also considered \W-densities $\tilde F$,
which we found could only exist in dimensions $\n=1,2$. The \W-density $\tilde
F$ generated an infinite set of gauge fields $\tilde h_{(n)}^{\mu_{1}
...\mu_{n}}$. In the case $\n=1$, these gauge fields transformed under local
${\rm Diff}_0(T^*\N) \sim \W _\infty$ transformations as
$$
\delta \tilde h_{(p)}= \sum_{m,n} \delta_{m+n,p+2}
\left[ (m-1) \lambda _{(m)}\partial
\tilde h_{(n)} -(n-1)\tilde h_{(n)}\partial \lambda _{(m)}
\right]\eqn\vogga$$
For $\n=2$,  we considered gauge fields with the transformation
$$\eqalign{
\delta\mathop{\tilde{h}}\nolimits^{\mu_{1}\mu_{2}...\mu_{p}}_{(p)}&=\sum
_{m,n}\delta_{m+n,p+2}\biggl[(m-1)\lambda^{(\mu_{1}\mu_{2}...}_{(m)}\partial
_{\nu}\mathop{\tilde{h}}\nolimits^{...\mu_{p})\nu}_{(n)}-(n-1)\mathop{\tilde
{h}}\nolimits^{\nu(\mu_{1}\mu_{2}...}_{(n)}\partial_{\nu}\lambda^{...\mu
_{p})}_{(m)}
\cr &
+{(m-1)(n-1)\over p-1}\partial_{\nu}\left\lbrace\lambda^{\nu(\mu
_{1}\mu_{2}...}_{(m)}\mathop{\tilde{h}}\nolimits^
{...\mu_{p})}_{(n)}-\mathop{\tilde
{h}}\nolimits^{\nu(\mu_{1}\mu_{2}...}_{(n)}\lambda^{...\mu_{p})}_{(m)}\right
\rbrace\biggr]
\cr}
\eqn\denvara
$$
Note that we could consider  this transformation for any dimension $\n$; in
particular, it reduces to \vogga\ if $\n=1$. However, for $\n>2$, the
corresponding generating function $\tilde F$ is never a \W-density, while for
$\n=2$  $\tilde F$ is not a \W-density for the  full group ${\rm
Diff}_0(T^*\N)$ but only for    the subgroup defined by the constraint
\xiszer, or equivalently, \conl. This formulation is redundant, in the sense
that there are more gauge fields than are needed, and it was shown that the
following constraint could be consistently imposed on the gauge fields:
 $$
\det \left ({\partial ^2   \tilde F (x,y)\over \partial y_\mu
\partial y_ \nu} \right)=-1
\eqn\deconcon$$
This constraint is preserved by the gauge transformations \denvara, and
imlies that
the linearised gauge fields are traceless.

For $\n=1$, we showed that gauge fields $\tilde h_{(n)}$ transforming as in
\vogga\ could be explicitly constructed from gauge fields $ h_{(n)}$
transforming as in \scaa, so that given any $\n=1$ \W-scalar $F(x,y)$, we
obtain a \W-density by writing
$$
\tilde F (x,y)= {\bf N} ^{-1} \left[ y \sqrt {2F(x,y)} \right]
\eqn\werwgra$$
For $\n=2$, we showed that the first few gauge fields $\tilde h_{(n)}$,
$n=2,3,4$, could be expressed in terms of \W-scalar gauge fields $ h_{(n)}$
(provided that the gauge fields satisfied
 the constraints generated by \deconcon)
and conjectured that there was such a construction for all $n$. The
reformulation in terms of the $ h_{(n)}$ involved many redundant gauge fields
(in the linearised theory, these are the traces of the $ h_{(n)}$) which could
be gauged away using \W-Weyl transformations.

The action for a single scalar field $\phi$ coupled to \W-gravity in either
one or two dimensions ($\n=1,2$) is then given by the integral of the
\W-density $\tilde F (x, \partial \phi)$ over $\N$,
 $$
S=\int d^{\n} x \sum_n
\tilde h^{\mu _1 \dots \mu
_n}_{(n)}(x)
S_{\mu _1 \dots \mu
_n}^{(n)}
 \eqn\cometdo$$
where
the currents $S_{\mu _1 \dots \mu
_n}^{(n)}$ are defined by
$$
S_{\mu _1 \dots \mu
_n}^{(n)}= {1 \over n} \partial_{\mu _1 }\phi\dots \partial_{\mu _n}\phi
\eqn\scurs$$
If $\n =2$, this remains invariant if the constraints generated by \deconcon\
are imposed on the gauge fields $\tilde h_{(n)}$, and it seems that the action
can then be reformulated in terms of \W-scalar gauge fields  $ h_{(n)}$.

So far we have restricted ourselves to the rather trivial case of a single
boson. However, for any matter current $S_{\mu \nu}$ that transforms under
diffeomorphisms in the same way as the free boson current ${1 \over 2} \gij
\dm \ffi \dn \fj$, \ie\ which transforms as a tensor, the action $ S=\int d^\n
x \, \tilde h^{\mu \nu }S_{\mu \nu} $ is invariant provided that $\tilde
h^{\mu \nu }$ transforms as a tensor density. In the same way, given any
matter system which can be used to construct a set of currents $S_{\mu _1
\dots \mu _n}^{(n)}$ which transform in the same way under \W-gravity
transformations as the single-boson currents \scurs, then the action \cometdo\
involving these new currents will be \W-invariant, provided that the gauge
fields $\tilde h$ transform as in  \vogga\ or \denvara. This immediately
gives  actions for  a large set of  matter systems; this will be discussed
further elsewhere.

Another important issue is the generalisation of these results to other
\W-algebras. As will be shown in [\wnprep], the gauge fields $\tilde h$ for
$\W_N$ gravity are  generated by a \W-density $\tilde F$ which, in addition to
the constraint \deconcon, satisfies a non-linear $(N+1)$'th order differential
constraint, which implies that only the gauge fields $\tilde h_{(2)}, \tilde
h_{(3)}, \dots \tilde h_{(N)}$ are independent. Whereas the constraint
\deconcon\ is related to self-dual geometry, the new  $(N+1)$'th order
differential constraint is similar to the type of constraint that arises in
the study of special geometry [\strom]. The  truncation to the \W-gravity
theory corresponding  to the algebra $\W_{\infty /2}$ is more straightforward:
it corresponds to setting to zero all of the gauge fields of odd spin,
$h_{(2n+1)}$.

One motivation for the study of \W-geometry is to try to understand finite
\W-transformations (as opposed to those with infinitesimal parameters) and the
moduli space for \W-gravity. The infinitesimal transformations for the scalar
field $\phi$ were derived from studying infinitesimal symplectic
diffeomorphisms and it follows that the large W-transformations of $\phi$ are
given by the action of large ${\rm Diff}_0 T^* \N$ transformations on $y_\mu =
\partial _\mu \phi$. The finite transformations of the gauge fields $h_{(n)}$
are given by requiring the invariance of the generating function $F(x,y)$,
while the finite transformations for the $\tilde h_{(n)}$ follow from
requiring the invariance of $\int \tilde F$, or from the construction of
$\tilde F$ in terms of $F$.  It seems natural to conjecture that the
transformations of the gauge fields can be defined to give invariance under
the full group of symplectic diffeomorphisms, as opposed to invariance under
the subgroup   generated by exponentiating infinitesimal ones, but this
remains to be proved.

The gauge-fixing of \W-gravity and the generalisation of the Liouville theory
that emerges in \W-conformal gauge were discussed in [\wanog]. Consider now
the moduli space $M_n$ for gauge fields $\tilde h_{(n)}$ subject to the
constraints generated by \deconcon\ [\wanog]. Linearising about a Euclidean
background $\tilde F = \half \tilde h_{(2)} ^{\mu \nu}y_{\mu \nu}$ and choosing
complex coordinates $z,\bar z$ on the Riemann surface $\N$ such that the
background is $\tilde F= y_z y_{\bar z}$,
 and using  the linearised transformations $\delta  \tilde h_{(n)}^{zz \dots
z}= \partial _{\bar z} \lambda _{(n)}^{zz \dots z}$, it follows by standard
arguments that the tangent space to the moduli space  $M_n$ at a  point
corresponding to the background configuration is the space of holomorphic
$n$-differentials, \ie\ the $n$-th rank symmetric tensors $\mu _{zz \dots z}$
with $n$ lower $z$ indices satisfying $\partial _{\bar z} \mu _{zz \dots z}=0$
[\wanog]. It follows from the Riemann-Roch theorem that the dimension of this
space on a genus-$g$ Riemann surface (the number of anti-ghost zero-modes) is
$dim(M_n)= (2n-1)(g-1)+k(n,g)$ where $k(n,g)$ is the number of solutions
$\kappa ^{zz \dots z}$ (with $n-1$  \lq $z$' indices) to $\partial _{\bar
z}\kappa ^{zz \dots z}=0$ (the number of ghost zero-modes). It would  be of
great interest to use information about the global structure of the symplectic
diffeomorphism group to learn more about the structure of these moduli spaces.

 \refout
\bye
\end